\documentclass[journal=jctc,manuscript=article]{achemso}

\usepackage[version=4]{mhchem} 
\usepackage{graphicx}
\usepackage{adjustbox}
\usepackage{makecell}
\usepackage{xr}

\usepackage[labelfont=bf, labelsep=space]{caption}
\usepackage{color,soul}
\usepackage{siunitx}

\graphicspath{{./Figures/}}

\usepackage{float} 

\usepackage{newtxmath}
\usepackage{upgreek}
\usepackage{bm}
\usepackage{multicol,multirow}
\usepackage{longtable}

\makeatletter
\newcommand*{\addFileDependency}[1]{
  \typeout{(#1)}
  \@addtofilelist{#1}
  \IfFileExists{#1}{}{\typeout{No file #1.}}
}
\makeatother

\newcommand*{\myexternaldocument}[1]{%
    \externaldocument{#1}%
    \addFileDependency{#1.tex}%
    \addFileDependency{#1.aux}%
}
\myexternaldocument{SI}

\author{Carmen Baiano}
\author{Jacopo Lupi}
\author{Vincenzo Barone}
\email{vincenzo.barone@sns.it}
\author{Nicola Tasinato}
\email{nicola.tasinato@sns.it}
\affiliation {Scuola Normale Superiore, Piazza dei Cavalieri 7, I-56126, Pisa, Italy}

\title[An \textsf{achemso} demo]
  {Gliding on ice in search of accurate and cost-effective computational methods for astrochemistry on grains: the puzzling case of the HCN isomerization}

\abbreviations{Astrochemistry,DFT,Benchmark}

\begin{document}

\begin{tocentry}
\centering
\includegraphics[width=7cm]{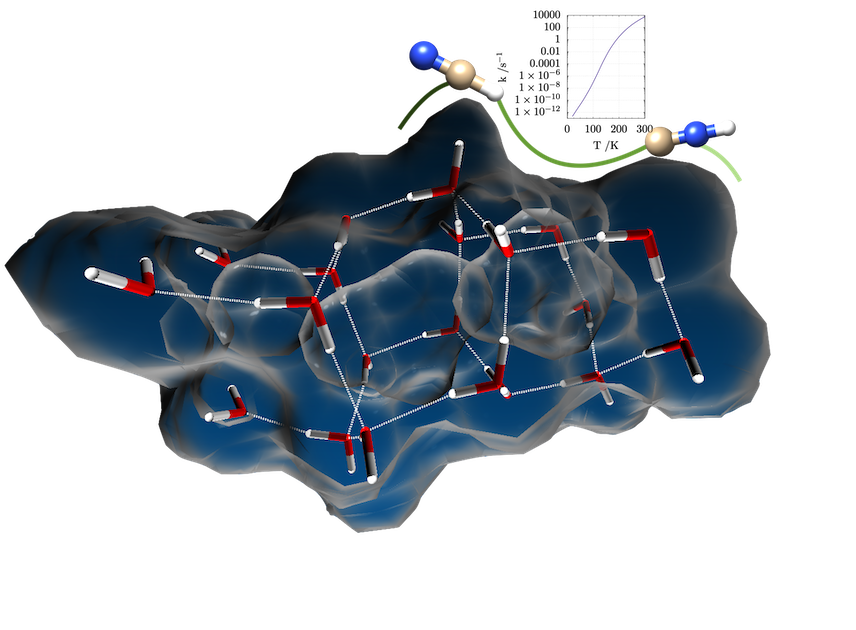}

\end{tocentry}

\begin{abstract}
The isomerization of hydrogen cyanide to hydrogen isocyanide on icy grain surfaces is investigated by an accurate composite method (jun-Cheap) rooted in the coupled cluster ansatz and by density functional approaches. After benchmarking density functional predictions of both geometries and reaction energies against jun-Cheap results for the relatively small model system \ce{HCN\bond{...}(H2O)2} the best performing DFT methods are selected. A large cluster containing 20 water molecules is then employed within a QM/QM$'$ approach to include a realistic environment mimicking the surface of icy grains. Our results indicate that four water molecules are directly involved in a proton relay mechanism, which strongly reduces the activation energy with respect to the direct hydrogen transfer occurring in the isolated molecule. Further extension of the size of the cluster up to 192 water molecules in the framework of a three-layer QM/QM$'$/MM model has a negligible effect on the energy barrier ruling the isomerization. Computation of reaction rates by transition state theory indicates that on icy surfaces the isomerization of HNC to HCN could occur quite easily even at low temperatures thanks to the reduced activation energy that can be effectively overcome by tunneling.
\end{abstract}

\section{1. Introduction}

In 2018, McGuire published a census of Interstellar, Circumstellar, Extragalactic, Protoplanetary Disks, and Exoplanetary Molecules\cite{McGuire_2018} including more than 200 molecules (containing from 2 to 70 atoms) and this number is steadily increasing thanks to the modern technologies of new observatory telescopes\cite{van_dishoeck_2017}.
The identification of many interstellar complex organic molecules (iCOMs) defeated the old and general idea that the interstellar medium (ISM) was an empty vial where chemical reactivity could not operate. 
Questions about the formation of iCOMs in such extreme conditions and the evolution of molecular complexity fueled the curiosity of astrochemists all over the world\cite{puzzarini-2020-Astro-challenges}.
While gas phase reactions seemed the obvious choice to explore the formation pathways of molecular systems in such  rarefied environments, the ubiquitous presence of dust and grains and the mismatch between some observations and the molecular abundances predicted by gas phase models have boosted the role of solid state chemistry\cite{Garrod_2008,Herbst-vanDishoeck-2009}.
Since the discovery of the catalytic role of grains for \ce{H2} formation\cite{Hollenbach1971,wakelam-2017}, astrochemists and physicists have struggled looking for gas-grain models that could provide a comprehensive picture of chemical processes in the ISM. 
At the low temperatures of molecular clouds (MC), molecules in the gas phase accrete icy mantles freezing out onto grain surfaces\cite{Allamandola-ices-1999,Burke-ices-2010} and leading to porous and amorphous icy surfaces\cite{watanabe-2008,Hama-2013,boogert-2015}, which can host local reactants triggering a molecular reactivity not feasible in the gas phase. The composition and morphological features make the simulation of these icy structures a great challenge in this field\cite{Cuppen2017,Burke-2010}.

The difficulty of performing experimental studies for systems capable of mimicking the harsh conditions of the ISM, calls for computational simulations of periodic surfaces and/or suitable model clusters able to take into the proper account the main structural features responsible for the chemistry at the interface\cite{Tasinato2018,Rimola-Modelling-2021}. This translates into the necessity of simulating extended systems, thus making the computational burden prohibitive for the accurate state-of-the-art methods developed for isolated molecules \cite{Barone2021}. 
Since water is the main component of polar icy mantles\cite{ehrenfreund&schutte-2000,Gibb&Tielens_2004}, a lot of efforts have been devoted to the investigation of the adsorption and formation of iCOMs on water clusters used to mimick interstellar ices.
The structures of H$_2$O clusters containing up to 22 atoms have been worked out from molecular dynamics simulations and made available in online databases\cite{Maheshwary-cluster-2001}. Some years ago, Rimola et al. studied iCOMs formation pathways on clusters including up to 33 water molecules obtained by combining two (H$_2$O)$_{18}$ clusters taken from the (010) surface of ice-XI\cite{Rimola-2010_Glycine} and removing three molecules to facilitate the construction of the final cluster\cite{Rimola-2018_formamide}.
Furthermore, attempts to include the structural modifications induced by UV and cosmic rays photo-processing have been made by means of small radical and ionized water clusters\cite{Rimola_2012-icePhotoprocessing}.
More recently, molecular dynamics has been used to model amorphous water ices\cite{Shimonishi_2018} and to simulate mixed CO/\ce{H2O} ices\cite{Zamirri_2018}. Adsorption energies on clusters of larger size have been evaluated by a two-layer our own N-layered integrated molecular orbital molecular mechanics (ONIOM) model, with the higher-level layer treated by means of density functional theory (DFT), and the lower-level one described through molecular mechanics (MM)\citep{Sameera2017,Sameera2021} or semiempirical quantum chemical methods\citep{Duflot2021}.

While coupled cluster theory including full treatment of single and double excitations together with perturbative estimation of triple excitations (CCSD(T)), possibly in conjunction with composite schemes to estimate the complete basis set (CBS) limit, is considered the gold-standard for accurate predictions \cite{Puzzarini2019}, the size of the systems to be dealt with in the case of ice-mediated chemistry makes density functional theory the only viable route in terms of accuracy to computational cost trade-off. As is well known, the reliability of DFT strongly depends on the specific system and properties at hand and on the choice of the density functional (DF) among an ever increasing number of possible formulations. In this respect, benchmark is a fundamental step for ranking the reliability of DFT model chemistries, also in connection with the computational cost, and hence it represents a very active field of research.

Concerning the specific topic of adsorption and reactivity of iCOMs on interstellar ice analogues, to the best of our knowledge, systematic benchmark studies are still lacking.  In this connection, Enrique-Romero et al.\cite{Romero-2019_radicals} performed a calibration analysis of radical-water interactions and activation energy for \ce{NH2 + HCO} and \ce{CH3 + HCO} reactions in the presence of one and two water molecules. They tested the accuracy of B3LYP and BHLYP functionals (both with and without dispersion corrections) in conjunction with the 6-311++G(2df,2pd) basis set taking CASPT2/cc-pVTZ and CCSD(T)/aug-cc-pVTZ levels of theory as reference. That analysis was focused on interaction and activation energies, while recent works have highlighted that reliable geometries are fundamental prerequisites for accurate thermochemistry and kinetics \cite{Barone2021}. In this respect, the B3LYP functional can be unable to predict correct structures for van der Waals complexes\cite{Spada2017b} and transition states\cite{puzzarini2020}. Furthermore, the use of CCSD(T)/triple-$\zeta$ energies cannot be recommended as a reference in benchmark studies because basis set truncation and lack of core-valence correlation limit the accuracy, thus introducing a bias in the reference values. This issue can be overcome by resorting to composite methods that aim at minimizing the errors relying on well-tested additive approximations \cite{Alessandrini-2019,Barone2021}.
\\
In this work, we assess the performances of several DFT model chemistries in evaluating the structural and energetic aspects of ice-mediated interstellar reactions employing the  HCN$\rightleftharpoons$HNC isomerization catalysed by water molecules as a paradigmatic process. On the one side, this can be considered a model for more complex reactions mediated by ice surfaces and, on the other side, the chosen system is small enough to allow the exploitation of state-of-the-art composite methods to generate accurate reference values for both geometries and reaction energies.
The HCN$\rightleftharpoons$HNC isomerization has been widely studied since the observed HNC/HCN ratio in the ISM can not be predicted on the basis of the proposed gas-phase mechanisms. Moreover, both HCN and HNC can be involved in the formation of amino acid precursors in the Strecker synthesis of glycine\cite{Woon-2001,Koch-2008_Aminoacetonitrile}. 
Gardebien et al. investigated the process for the isolated molecule and with explicit inclusion of two to four water molecules\cite{Gardebien-2003} finding that the most favourable mechanism consists of a one-step path involving a proton relay mediated by the water cluster. Koch et al. employed a more realistic model including seven additional water molecules to simulate the local environment of the icy surface and employing the polarizable continuum model (PCM) to account for bulk effects\cite{Koch-2007}.
According to the available data, the water cluster acts as a catalyst lowering the energy barrier with respect to the gas-phase, an effect that  progressively smooths increasing the number of H$_2$O molecules. Intermolecular proton transfer drives both the interaction of HCN and HNC with the ice surface and the isomerization process. This represents the most common mechanism through which molecules adsorb and react on ISM polar ices.\\
On these grounds, we decided to perform a detailed study of the HCN$\rightleftharpoons$HNC isomerization by state-of-the-art quantum chemical methods and realistic cluster models. The work is organized as follows: the computational methods are described in Section 2, while the outcomes of the benchmark are detailed in Section 3 concerning both geometries and energies, thus leading to the identification of the best performing DFT model chemistries in terms of the trade-off between accuracy and computational cost. Despite the fact that the benchmark is carried out on a simplified model, the outcomes are expected to be of general validity, especially with respect to the relative performances of the tested methods which can then be transferred to larger H$_2$O clusters. With this in mind, at the end of Section 3, the best performing methods are employed to simulate the HCN$\rightleftharpoons$HNC isomerization catalyzed by a cluster of twenty water molecules, then further embedded in a 172 water slab described through MM. Finally, reaction rates are computed in the framework of the transition state theory (TST) including tunneling.

\section{2. Computational Methodology}
 
For the benchmark study, we selected 10 DFs belonging to different families: two hybrids (B3LYP, BHLYP)\cite{Becke-1993,LYP_1988,VWN_1980}, a long-range corrected DF ($\omega$B97X-D)\cite{Chai-wB97XD-2008}, three meta-hybrids (PW6B95\cite{Zhao-PW6B95-2005}, BMK\cite{Boese_2004-BMK} and M06-2X\cite{Zhao&Truhlar-M062X-2007}), one meta-NGA (MN15\cite{Yu-MN15-2016}), the B2PLYP\cite{Grimme-2006-B2PLYP} and the two spin-component-scaled (DSD-PBEP86 and revDSD-PBEP86)\cite{Kozuch-DSD-2011,Kozuch-revDSD-2013} double hybrids. To test the accuracy to computational cost trade-off, for each functional six basis sets have been considered. In particular, we selected the Dunning's aug-cc-pV$n$Z basis sets ($n = $ D, T) \cite{Dunning1989,Kendall1992} as well as the corresponding jun- and jul- modifications from the Truhlar's calendar family.\cite{Truhlar-calendar-2011}  
All the DFT calculations include empirical dispersion corrections according to the DFT-D3 scheme proposed by Grimme\cite{Grimme-D3-2010} with the Becke-Johnson damping function (BJ)\cite{Grimme-BJ-2011,Smith-2016}, which are fundamental for the correct prediction of van der Waals complexes\cite{Burns2011,Klimes2012,Tasinato2015}, transition states\cite{Goerigk2011} and surface processes\cite{DellePiane2013,Tasinato2015b}.
Accurate reference geometries and energies for the benchmark were generated by using the Cheap composite scheme (ChS) \cite{Puzzarini-2011-ChS,Puzzarini-2013-ChS2} and its recent jun-Cheap revision (jun-ChS) \cite{Barone2021,Alessandrini-2019}, with the latter appearing the best option because of the increased reliability for non-covalent interactions and the better description of the water dimer structure. Indeed, for \ce{(H2O)2}, ChS and jun-ChS geometries were first compared to highly accurate CCSD(T)-F12b/CBS+fT+fQ+CV+REL+DBOC values.\cite{Lane-2013} The results, reported in Table S1 of the Supporting Information (SI), show that bond lengths and valence angles are reproduced very accurately, with maximum errors of -0.003 \r{A} and -0.2 \textdegree, while there is a deviation of 3 \textdegree for the angle defining the orientation of the $C_2$ axis of the acceptor water molecule with respect to the O$-$O axis. 
On the basis of the reliable geometry delivered by jun-ChS, this method was used as reference for both equilibrium geometries and electronic energies.\\ 
Preliminary B3LYP-D3/aug-cc-pVTZ computations of the HCN$\rightleftharpoons$HNC reactive PES were refined at the jun-ChS level. The nature of the identified stationary points (minima or saddle points) was checked through frequency calculations performed at each level of theory.
All calculations have been carried out with the Gaussian software \citep{g16}, except the geometry optimizations at the ChS and jun-ChS levels, which have been performed using the CFOUR package \citep{cfour,cfour-new}. Since revDSD-PBEP86 is not among the Gaussian built-in functionals, it has been defined by setting proper IOP flags on top of the DSD-PBEP86 functional. \\
Full geometry optimizations were performed for the complexes containing 2 to 4 \ce{H2O} molecules, whereas for the 20 water model cut from the ice XI (010) surface, 8 molecules belonging to the cluster edge (see Figure S1 of the SI) were kept frozen at their positions in the crystal in order to prevent geometrical distortions causing a non-physical breakdown of the crystalline pattern. The best-performing methods were employed within a QM/QM$'$ strategy for simulating the HCN$\rightleftharpoons$HNC isomerization on this cluster in order to evaluate the catalytic effect of the ice surface. For the purpose, we employed the ONIOM method \citep{Vreven2006} treating the reaction center (i.e., the adsorbate and four water molecules) at a higher level of theory (i.e. a double-hybrid DF or even jun-ChS), whereas a less computationally-demanding method (i.e. a meta-hybrid DF) was used for the remaining molecules of the cluster. A much larger cluster containing 192 water molecules was also investigated by means of a three-layer (QM/QM$'$/MM) ONIOM approach enforcing the so-called mechanical embedding and employing the Amber force field\cite{Cornell1995}. In this case, the structural degrees of freedom of the adsorbate and the first 20 \ce{H2O} molecules were optimized while freezing the coordinates of the remaining 172 waters to those of the regular (010) surface of ice XI. Test computations with the more refined electrostatic embedding showed negligible differences on the relative energies.

Rate constants were computed solving the multi-well one-dimensional master equation using the chemically significant eigenvalues (CSEs) method \citep{georgievskii2013reformulation}. Rate coefficients were determined using conventional transition state theory (TST) within the rigid-rotor harmonic-oscillator (RRHO) approximation \citep{TS_Truhlar}, also incorporating tunneling and non-classical reflection effects by means of the Eckart model \citep{eckart1930penetration}. 
The rates evaluated at different temperatures were fitted by a simple Arrhenius Equation or by the three-parameter modified Arrhenius equation proposed by Kooij\citep{kooij,laidler96}:
\begin{equation}
\label{eq:kooij}
    k(T)=A\left(\frac{T}{300}\right)^n\exp\left(-\frac{E_a}{RT}\right)
\end{equation}
where $A$, $n$, and $E_a$ are the fitting parameters, $R$ is the universal gas constant, and the limiting Arrhenius behaviour is recovered when $n=0$. All the kinetic computations were performed with the MESS code. \citep{georgievskii2013reformulation}.

\section{3. Results and discussion}

As widely discussed in the Introduction, the reliable modelling of interstellar ices is an extremely complex task, requiring the assessment of DFT methods for geometry and energy predictions that offer the proper balance between accuracy and computational burden. The lack of systematic studies addressing this issue for solid-state astrochemical processes calls for a dedicated benchmark.
While small-size clusters cannot be fully representative of an extended substrate, the interaction of small molecules with water ice surfaces is generally guided by hydrogen bonds between the polar functional groups of the molecule and the exposed H and O atoms of the ice surface, which are already present in the smallest cluster models. Therefore, while the thermochemistry computed by using clusters composed of a small number of \ce{H2O} molecules is not representative of real icy-grain chemistry, the outcomes of the benchmark are safely transferable to larger clusters. 
In the following subsections we report the results of our benchmark study, concerning first geometries and then reaction and activation energies. 
Finally, to scale-up to a more realistic water ice model, we report a full characterization of the PES of the HCN$\rightleftharpoons$HNC isomerization on clusters composed by either 20 or 192 \ce{H2O} molecules.

\subsection{3.1. The geometry snow-board}

\begin{figure}[!ht]

    \centering
    \includegraphics[width=12cm]{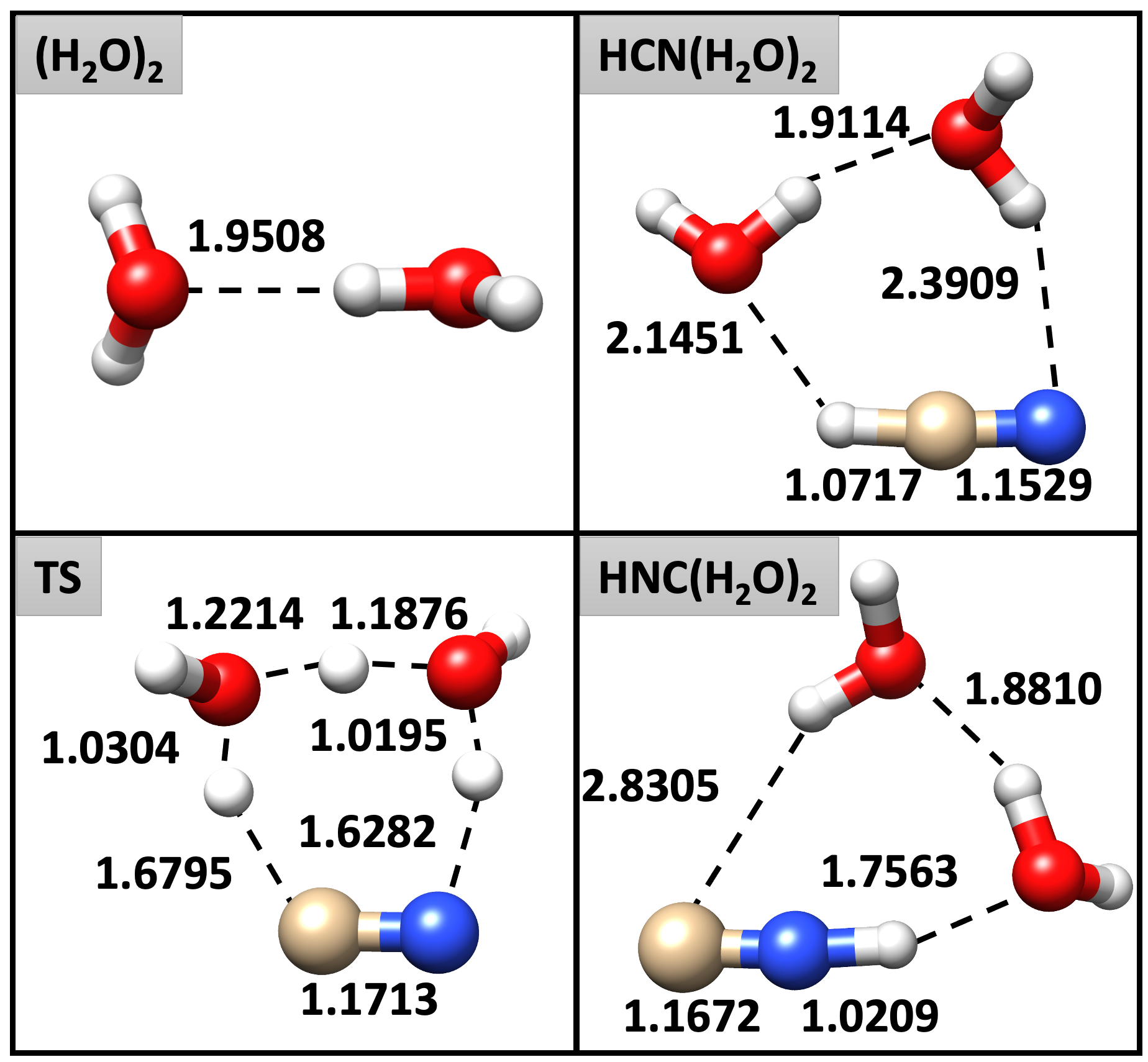}
    \caption{Stationary points on the reactive PES of the HCN$\rightleftharpoons$HNC isomerization catalysed by two water molecules. Representative bond lengths (\r{A}) obtained at jun-ChS level are reported.}
    \label{fig:struc_PES}

\end{figure}

The HCN$\rightarrow$HNC isomerization is an endothermic process involving a high activation energy and the jun-ChS results are close to the current best estimates \cite{zheng2009} for both the reaction (61.6 vs. 63.8 kJ/mol) and activation (198.5 vs. 201.1 kJ/mol) energy. Addition of two water molecules leads to the formation of a hydrogen-bonded van der Waals adduct featuring the interactions between the H atom of HCN and the oxygen of one water molecule and between the N atom and one hydrogen of the second water molecule. Then, the reaction proceeds through a transition state for the \ce{(H2O)2}-mediated proton transfer reaching, in this way, a post-reactive complex in which carbon is engaged in a weak H-bond with a hydrogen of the first \ce{H2O} molecule, while the H atom of HNC interacts with the oxygen of the second water molecule. The structures of all the stationary points ruling the reactive PES are sketched in Figure \ref{fig:struc_PES} together with selected geometrical parameters obtained at the jun-ChS level.

\begin{figure}[h]
    \centering
    \includegraphics[width=14cm]{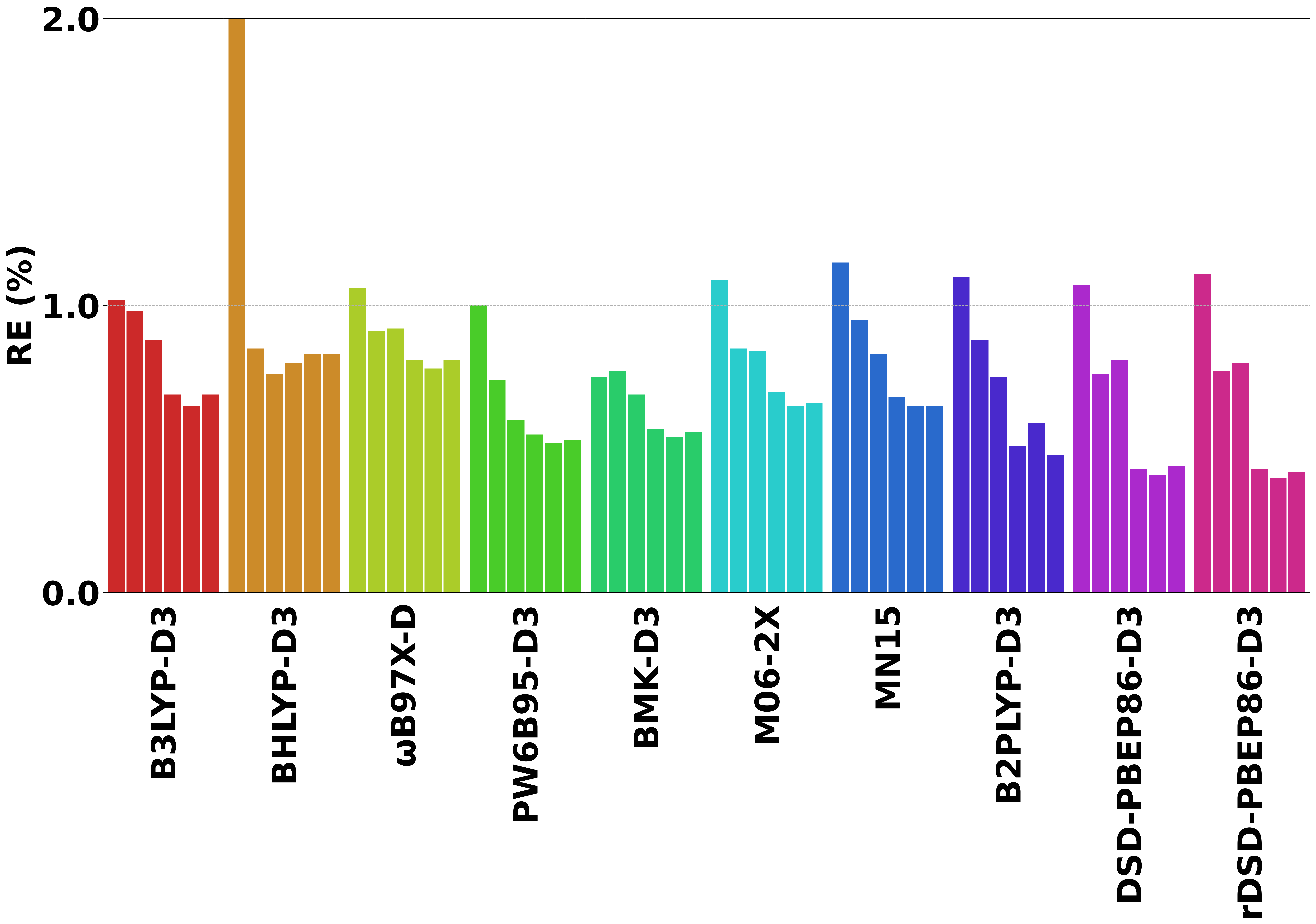}
    \caption{Total REs (\%) of the geometries of the species on the PES of the HCN$\rightleftharpoons$HNC isomerization assisted by two water molecules for the investigated DFT methods with respect to jun-ChS reference values. For each functional, the different basis sets are reported in the following order: jun-DZ, jul-DZ, aug-DZ, jun-TZ, jul-TZ and aug-TZ}
    \label{fig:RE-TOT}
\end{figure}

The accuracy of the considered DFT model chemistries has been evaluated with respect to jun-ChS values and the overall mean absolute errors (MAEs) and mean absolute relative errors (REs) have been evaluated over all the bond lengths, valence and dihedral angles of the species involved in the PES. The full list of data can be found in Table S2 and Figure \ref{fig:MAEandRE} of the SI. As a rule of thumb (with some exceptions for dihedral angles), triple-$\zeta$ basis sets show smaller errors than the corresponding double-$\zeta$ ones, with the improvement being less pronounced along the jun-, jul- and aug- series. In general, the tested hybrid and meta-hybrid DFs on the one side, and the double-hybrids on the other, give similar trends for the MAEs, with the notable exception of BHLYP-D3 in conjunction with the jul-cc-pVDZ basis set, that strongly overshoots and the $\omega$B97X-D functional that shows larger deviations from the jun-ChS reference values, especially for valence and dihedral angles. In the case of the BHLYP-D3/jul-cc-pVDZ model, MAEs as large as 0.09 \r{A}, 6\textdegree and 10\textdegree were observed for bond lengths, valence and dihedral angles, respectively. These results are related to the inability of reproducing a tight structure for the post-reactive complex. Specifically, one H-bond in CNH--(H$_{2}$O)$_{2}$ (see Figure \ref{fig:struc_PES}) is broken and the product collapses into an open structure. All in all, it can be observed that the most promising (meta-)hybrid DFs are PW6B95-D3, BMK-D3, M06-2X and MN15 coupled to triple-$\zeta$ basis sets (or, at least, the jul-cc-pVDZ one). Concerning the double-hybrid functionals, the best structural predictions are delivered by DSD-PBEP86-D3 and revDSD-PBEP86-D3 that show comparable accuracy.   
In order to have a clearer picture of the performance of the different model chemistries in the prediction of the geometries involved in the HCN$\rightleftharpoons$HNC isomerization assisted by two water molecules, Figure \ref{fig:RE-TOT} reports the overall REs of each method, evaluated by averaging the REs of the geometrical parameters of all the species on the reactive PES. Inspection of this figure reveals that, among the (meta-)hybrid DFs the best results for double-$\zeta$ basis sets are delivered by PW6B95-D3 and BMK-D3. In particular, PW6B95-D3/jul-cc-pVDZ, BMK-D3/aug-cc-pVDZ and PW6B95-D3/aug-cc-pVDZ score REs in the 0.60\% - 0.74\% range. The PW6B95-D3 and BMK-D3 DFs are the best performers also in conjunction with triple-$\zeta$ basis sets showing REs around 0.55\%. Concerning the double-hybrid functionals, it is apparent that their use in conjunction with a double-$\zeta$ basis set does not justify the computational overload in comparison with hybrid functionals; however, both DSD-PBEP86-D3 and its revision predict improved geometries when employed in conjuncion with triple-$\zeta$ basis sets, reaching a RE of only 0.4\% for the jul-cc-pVTZ basis set. In passing, it is interesting to point out that, these functionals have also demonstrated to be excellent performers in predicting structural and spectroscopic properties of gas-phase molecules \cite{Barone2020,Ceselin2021}.

\subsection{3.2. Skiing on adsorption, reaction and activation energies}

The functional/basis set combinations with the optimal accuracy/cost trade-off for geometry predictions have been identified in the previous section.
Reactivity studies require the calculation of accurate formation and activation energies for the subsequent kinetic analysis.
For this reason, some of the DFT methods delivering the best geometrical predictions have been selected and their accuracy for computing adsorption, activation and reaction (electronic) energies explored using again jun-ChS results as references. 
In a first step, the impact of the geometry on the energetics has been assessed, by evaluating jun-ChS electronic energies for the different DFT structures. In a second step, the formation energies stemming from full DFT computations (for both geometries and energies) have been analysed. 

\begin{figure}[!ht]

    \centering
    \includegraphics[width=8cm]{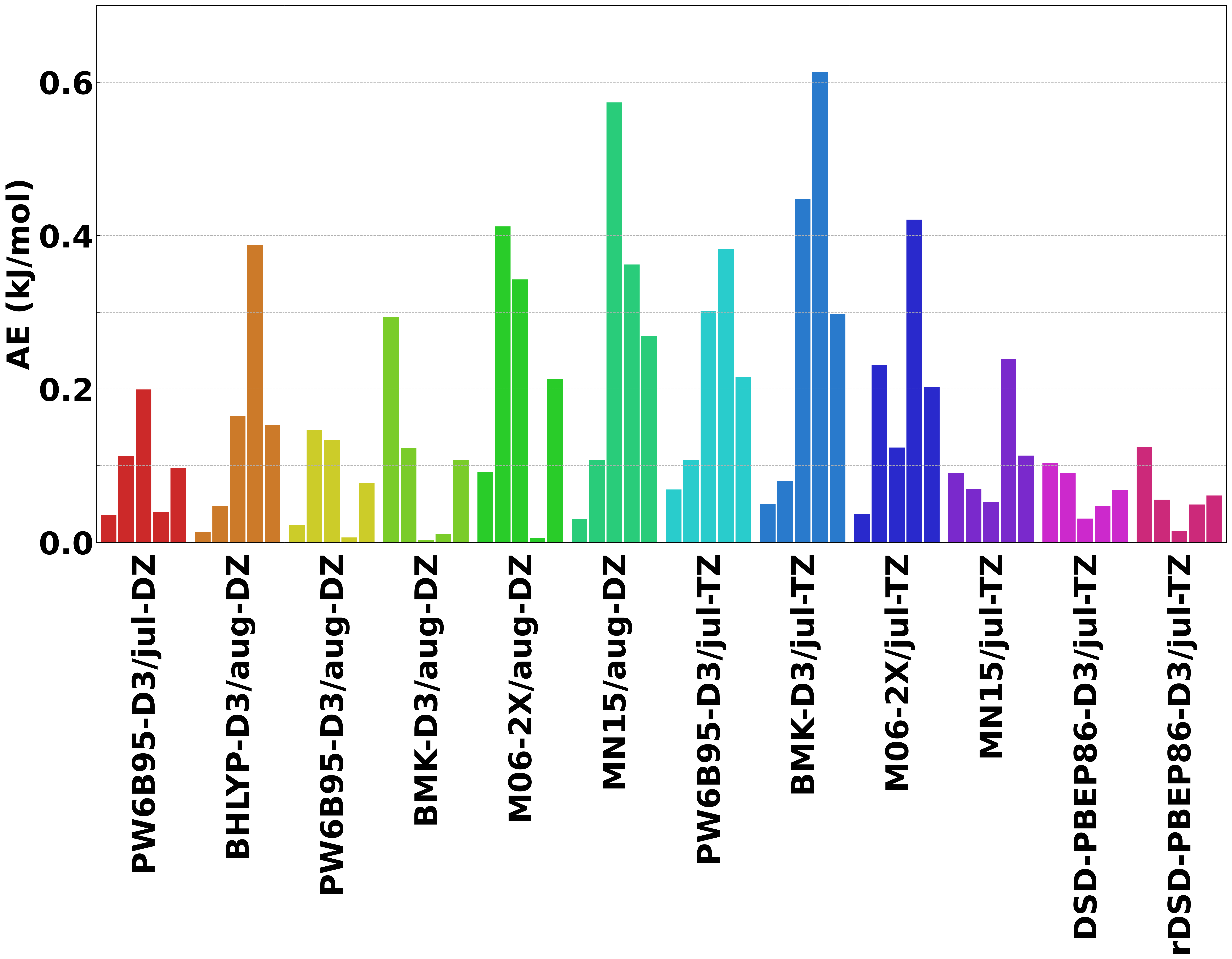}
    \caption{Error analysis for jun-ChS formation energy (kJ/mol) obtained on top of DFT geometries in comparison with full (both energies and geometries) jun-ChS results. Each color corresponds to a DFT model chemistry and collects absolute errors for the formation energy of each species along the PES with respect to isolated reactants: 1. pre-reactive complex; 2. transition state; 3. post-reactive complex; 4. products; 5. MAE over all of the steps along the PES.}
    \label{fig:MAE-deltaE}

\end{figure}

Electronic energies obtained at the jun-ChS level on top of selected DFT geometries are reported in Table \ref{table:Formation_ene}, while the corresponding error analysis is presented in in Figure \ref{fig:MAE-deltaE} and in Table S3 of the SI. \\

\begin{table}
\caption{jun-ChS formation energies (kJ/mol) with respect to isolated HCN and \ce{(H2O)2} for each species along the HCN/\ce{HNC\bond{...}(H2O)2} isomerization PES evaluated on top of DFT geometries.}
\label{table:Formation_ene} 
\begin{tabular}{lcccc}
\hline
\textbf{Level of theory for geometry}                      & \textbf{\ce{HCN\bond{...}(H2O)2}} & \textbf{TS} & \textbf{\ce{CNH\bond{...}(H2O)2}} & \textbf{\ce{HNC + (H2O)2}} \\
\hline
\textbf{PW6B95-D3/jul-DZ}      & -33.38        & 99.37       & 15.23         & 62.31              \\
\textbf{BHLYP-D3/aug-DZ}       & -33.40        & 99.21       & 14.87         & 61.88              \\
\textbf{PW6B95-D3/aug-DZ}      & -33.44        & 99.40       & 15.17         & 62.28              \\
\textbf{BMK-D3/aug-DZ}         & -33.12        & 99.13       & 15.04         & 62.26              \\
\textbf{M06-2X/aug-DZ}         & -33.32        & 99.67       & 15.38         & 62.28              \\
\textbf{MN15/aug-DZ}           & -33.39        & 99.36       & 15.61         & 62.63              \\
\textbf{PW6B95-D3/jul-TZ}      & -33.49        & 99.15       & 14.73         & 61.89              \\
\textbf{BMK-D3/jul-TZ}         & -33.47        & 99.18       & 14.59         & 61.66              \\
\textbf{M06-2X/jul-TZ}         & -33.38        & 99.49       & 14.91         & 61.85              \\
\textbf{MN15/jul-TZ}           & -33.51        & 99.33       & 14.98         & 62.03              \\
\textbf{DSD-PBEP86-D3/jul-TZ}  & -33.52        & 99.35       & 15.06         & 62.22              \\
\textbf{revDSD-PBEP86-D3/jul-TZ} & -33.54        & 99.31       & 15.02         & 62.22              \\
\hline
\textbf{jun-ChS}           & -33.42        & 99.26       & 15.03         & 62.27          \\
\hline
\end{tabular}
\end{table}

\begin{table}
\caption{DFT formation energies (kJ/mol) with respect to isolated HCN and water dimer (H$_{2}$O)$_{2}$ for each species along the \ce{HCN\bond{...}(H2O)2} isomerization PES.}
\label{table:Formation_ene-DFTonDFTgeom}
\begin{tabular}{lcccc}
\hline
\textbf{Level of theory}$^a$                      & \textbf{\ce{HCN\bond{...}(H2O)2}} & \textbf{TS} & \textbf{\ce{CNH\bond{...}(H2O)2}} & \textbf{\ce{HNC + (H2O)2}} \\
\hline

\textbf{PW6B95-D3/jul-DZ}      & -34.12 & 86.96 & 8.34  & 56.74            \\
\textbf{BHLYP-D3/aug-DZ}       & -38.42 & 89.54 & -1.35 & 51.12            \\
\textbf{PW6B95-D3/aug-DZ}      & -33.99 & 88.45 & 7.70  & 56.03            \\
\textbf{BMK-D3/aug-DZ}         & -36.30 & 81.76 & -6.60 & 42.58            \\
\textbf{M06-2X/aug-DZ}         & -37.46 & 70.46 & 1.14  & 52.78            \\
\textbf{MN15/aug-DZ}           & -36.67 & 76.94 & -2.82 & 47.31            \\
\textbf{PW6B95-D3/jul-TZ}      & -32.67 & 94.10 & 11.31 & 56.81            \\
\textbf{BMK-D3/jul-TZ}         & -35.97 & 88.16 & 0.11  & 48.23            \\
\textbf{M06-2X/jul-TZ}         & -37.16 & 76.00 & 4.64  & 54.06            \\
\textbf{MN15/jul-TZ}           & -35.26 & 85.53 & 1.92  & 48.88            \\
\textbf{DSD-PBEP86-D3/jul-TZ}  & -35.08 & 88.86 & 14.86 & 64.93            \\
\textbf{revDSD-PBEP86-D3/jul-TZ} & -33.97 & 93.50 & 16.21 & 64.85          \\
\hline
\textbf{jun-ChS}                  & -33.42 & 99.26 & 15.03 & 62.27 \\          
\hline
\end{tabular}
$^a$ For both energy and geometry.
\end{table} 

It is quite apparent that the energetic results obtained employing geometries optimized with all the tested methods are in remarkable agreement with the jun-ChS reference, with deviations smaller than 0.6 kJ/mol, even though some of them provide an unbalanced description of the different elementary processes. For example, the MN15/aug-cc-pVDZ and BMK-D3/jul-cc-pVTZ models yield excellent predictions of both the interaction energy of hydrogen cyanide with \ce{(H2O)2} and the transition state energy, with errors around 0.05 and 0.1 kJ/mol, respectively; however, the computed HNC formation energy (at the BMK-D3 level) and its interaction energy with the water dimer (at the MN15 level) show significantly larger errors.  
Among the (meta-)hybrid functionals, the best and most consistent energetic description is given by PW6B95-D3 in conjunction with jul- or aug-cc-pVDZ basis sets, which reaches an overall MAE (evaluated by considering the relative electronic energies of all the stationary points ruling the PES) close to 0.1 kJ/mol and a maximum deviation of 0.2 kJ/mol. 

\begin{figure}[!ht]

    \centering
    \includegraphics[width=8cm]{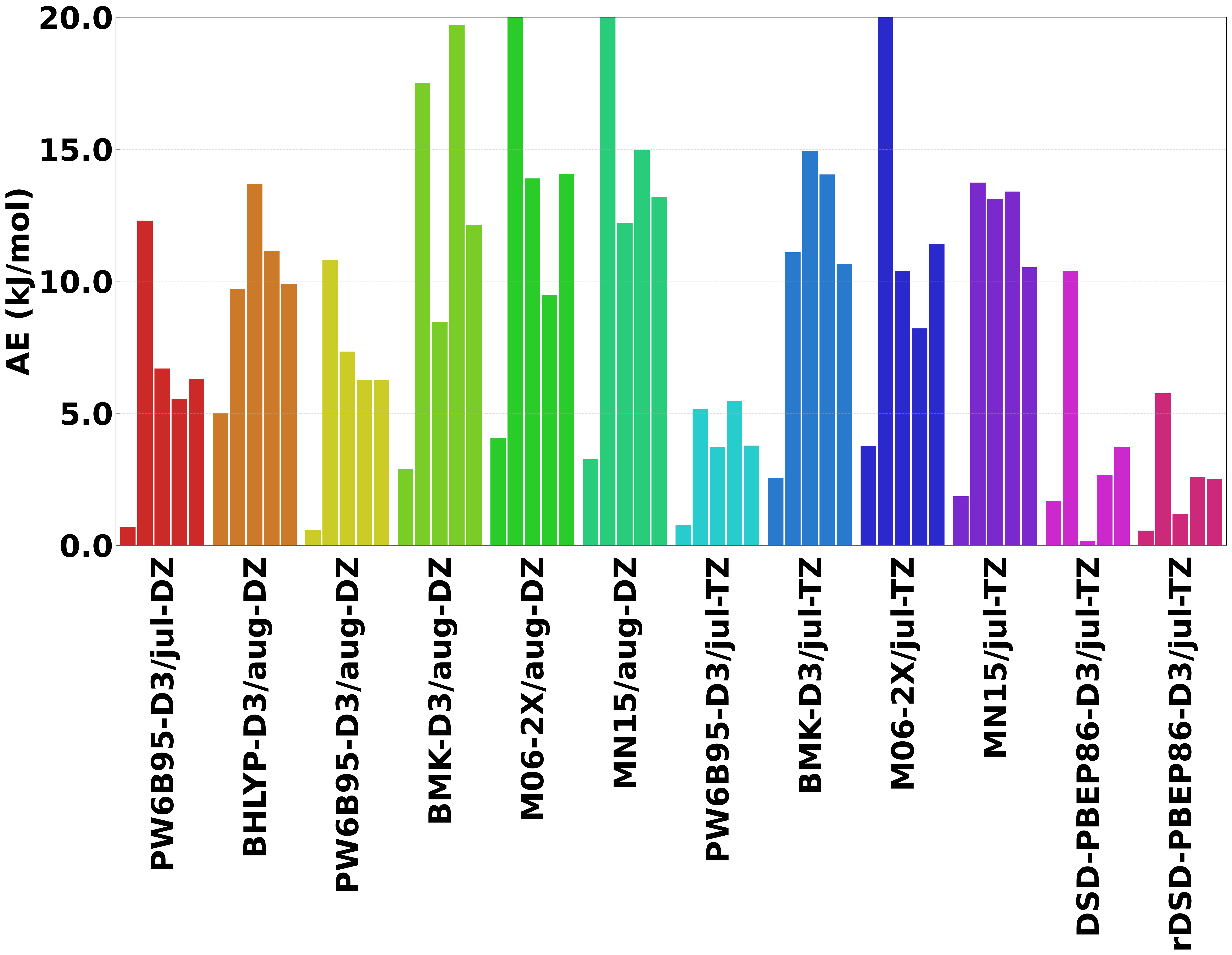}
    \caption{Error analysis for DFT formation energies (kJ/mol) in comparison with jun-ChS values. Each color corresponds to a DFT model chemistry (used for both geometry and energy) and collects absolute errors for the formation energy of each species along the PES with respect to isolated reactants: 1. pre-reactive complex; 2. transition state; 3. post-reactive complex; 4. products; 5. MAE over all of the steps along the PES.}
    \label{fig:MAE-deltaE-DFTE-on-DFTg}
\end{figure}

Moving to the double-hybrid DFs, the DSD-PBEP86-D3 and revDSD-PBEP86-D3 models in conjunction with the jul-cc-pVTZ basis set yield excellent performances, scoring a MAE of about 0.06 kJ/mol and reproducing the formation energies of all the elementary steps with a maximum deviation of 0.12 kJ/mol for the pre-reactive complex at the revDSD-PBEP86-D3/jul-cc-pVTZ level. \\
The relative electronic energies of all the stationary points fully evaluated at different DFT levels (i.e. energies and geometries) are collected in Table \ref{table:Formation_ene-DFTonDFTgeom} and the MADs from the jun-ChS computations are shown in Figure \ref{fig:MAE-deltaE-DFTE-on-DFTg}. In general terms, the results mirror those obtained for jun-ChS energies evaluated on top of DFT geometries, with the only difference being the much larger deviations, which now span the 5 - 29 kJ/mol range. Furthermore, the relative stability of \ce{CNH\bond{...}(H2O)2} is always strongly underestimated (becoming even negative with BHLYP-D3, BMK-D3 and MN15 functionals) except at the PW6B95-D3 and, especially, DSD-PBEP86-D3 and revDSD-PBEP86-D3 levels in conjunction with the jul-cc-pVTZ basis set.  All the (meta-)hybrid DFs show MAEs larger than 10 kJ/mol, with the exception of PW6B95-D3, which is the only functional that reaches a MAE around 6 kJ/mol in conjunction with the jul- and aug-cc-pVDZ basis set and of 3.8 kJ/mol employing the jul-cc-pVTZ basis. The DSD-PBEP86-D3 and revDSD-PBEP86-D3 functionals confirm their good performances in conjunction with the jun-cc-pVTZ basis set, with MAD around 3 kJ/mol and maximum deviations of 10.4 kJ/mol. Hence, the model chemistries with the optimal accuracy for structural parameters are also the best choices for thermochemistry.\\
These results confirm the conclusions of recent benchmarks about the quality of PW6B95-D3/jul-cc-pVDZ and DSD-PBEP86-D3/jul-cc-pVTZ models for geometries, vibrational frequencies and other spectroscopic parameters. \cite{Barone2020,Ceselin2021} Noted is that core-valence correlation has not been included for double hybrid functionals because it was not taken into account in their original parametrization and its contribution is anyway within the expected error bar at least for molecular systems containing only hydrogen and second-row atoms (see SI for CV contributions in jun-ChS results). Furthermore, some test computations performed with quadruple-$\zeta$ basis sets showed that complete basis set (CBS) extrapolation has a negligible effect on all the trends discussed above. For example, the relative electronic energies of the stationary points obtained by using the DSD-PBEP86-D3 functional in conjunction with the aug-cc-pVQZ basis set ($\Delta E = $ -35.04, 89.51, 14.83 and 64.78 kJ/mol) differ from the counterparts obtained employing the jul-cc-pVTZ basis set by 0.65 kJ/mol at most (for the TS). Finally, although triple-$\zeta$ basis sets possibly deliver more robust results for hybrid functionals, this computational level will be used in the following only to describe small environmental effects in the framework of QM/QM$'$ computations where the increased computational cost with respect to double-$\zeta$ results is not justified, in our opinion, by the marginally improved robustness. 
 
\subsection{Scaling-up toward extended systems: best performers at work}

The benchmark performed for both geometries and energies permits the identification of the best candidates for setting up a QM/QM$'$ ONIOM strategy for the study of the HCN$\rightleftharpoons$HNC isomerization on large clusters capable of providing a more realistic modelling of the icy-grain and of the molecule-surface interactions. 

\begin{figure}[!ht]

    \centering
    \includegraphics[width=\textwidth]{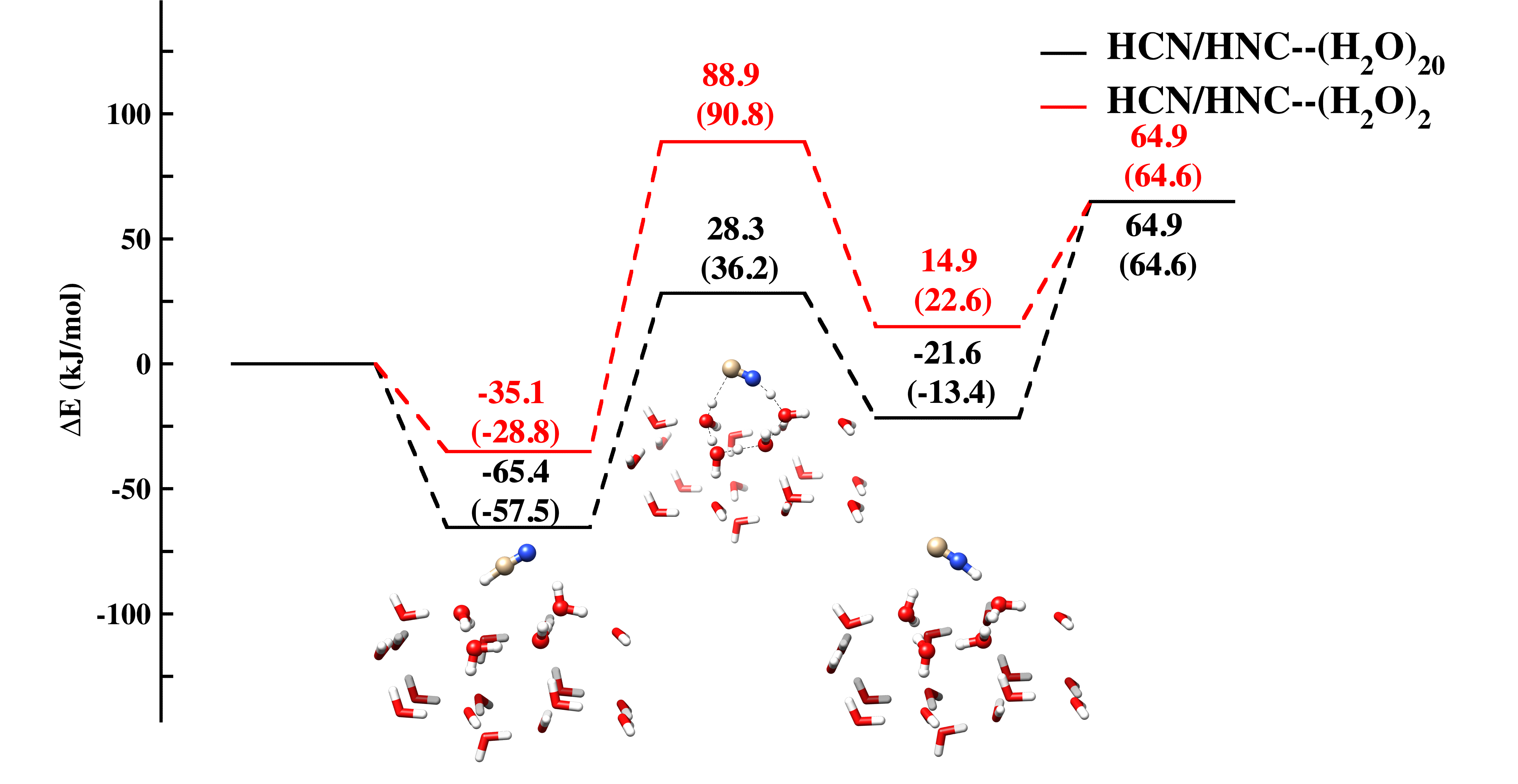}
    \caption{Potential energy profile for HCN$\rightleftharpoons$HNC isomerization mediated by the \ce{(H2O)20} cluster and the \ce{(H2O)2} dimer. Red lines refer to the HCN isomerization catalysed by \ce{(H2O)2} and both geometries and $\Delta$E have been computed at the DSD-PBEP86-D3/jul-cc-pVDZ level. Black lines refer to the ONIOM results for the reaction catalysed by  \ce{(H2O)20}. The ball and stick representation is used for atoms of the highest QM level (DSD-PBEP86-D3/jul-cc-pVDZ) while the tube representation is used for the atoms belonging to the QM$'$ (PW6B95-D3/jul-cc-pVDZ) portion. $\Delta$E corrected for ZPVE are reported in parenthesis with ZPVEs calculated at the same level of theory as the corresponding energies and geometries. }
    \label{fig:PES_H2O20-ONIOM}

\end{figure}

At first, a cluster composed by 20 water molecules (shown in Figure \ref{fig:PES_H2O20-ONIOM} and Figure S1 of the SI) has been used, in which the pattern of exposed water molecules is suitable for a H-relay mechanism mediated by four water molecules. It should be noted that in Ref.\cite{Koch-2007} a proton-relay mechanism mediated by three water molecules, in turn solvated by seven additional waters, was used. In the present work, the four \ce{H2O} molecules involved in the hydrogen transfer and the adsorbed species have been considered as the reaction center of the process under study, hence they constitute the higher-level QM portion of the system. Following the outcomes of the benchmark study, the DSD-PBEP86-D3 functional in conjunction with the jul-cc-pVTZ basis set has been used for the purpose, while the remaining part of the cluster, treated at a lower QM$'$ level, has been described by the PW6B95-D3 DF in conjunction with the jul-cc-pVDZ basis set.
The energetic profile of the HCN$\rightleftharpoons$HNC isomerization occurring on the \ce{(H2O)20} cluster is reported in Figure \ref{fig:PES_H2O20-ONIOM} where it is also compared with that for the \ce{(H2O)2}-mediated process. Going from the process assisted by two waters to that assisted by four water molecules in the \ce{(H2O)20} cluster lowers the energy of all the species present in the reactive PES. The most remarkable effect is the reduction of the energy barrier ruling the isomerization when considering the 20 water cluster in place of just two water molecules involved in the simplest possible relay mechanism. 

\begin{table}

\caption{Relative ground state energies (kJ/mol) with respect to HNC-\ce{(H2O)$_n$} post reactive complex and comparison with the results of Ref. \cite{Koch-2007}. Both the total number of water molecules ($n$) and the number of water molecules directly involved in the relay mechanism ($n_R$) are indicated. All values include ZPVEs.}
\label{table:SI_formationenergies}

\begin{tabular}{p{4.8cm}lcccc}
\hline
& \textbf{total \ce{H2O}} ($n$) &  \textbf{relay \ce{H2O} ($n_R$)} & \textbf{TS}$^a$   & \textbf{\ce{HCN\bond{...}(H2O)_n}}$^a$ \\
\hline

\multirow{3}{*}{\textbf{Ref. \cite{Koch-2007}}} & 2 & 2    & 74.1 & -42.3  \\
                                    & 3 & 3            & 43.9 (-30.2) & -41.4 (-0.9) \\
                                    & 10 & 3           & 13.8 (-60.3) & -41.4 (-0.9) \\
\hline

\multirow{2}{*}{\textbf{B3LYP}$^b$}  & 2 & 2           & 73.8         & -39.8        \\
                                     & 3 & 3           & 44.1 (-29.7) & -41.7 (-1.9) \\
\hline
\multirow{2}{*}{\textbf{PW6B95-D3}$^c$} & 2 & 2              & 70.9             & -43.6            \\
                                     & 3 & 3              & 52.4 (-18.5)     & -42.0 (-1.6)     \\
                                     & 4 & 4              & 49.3 (-21.6)     & -42.7 (-0.9)     \\
                                     & 192 & 4            & 36.6$^d$ (-33.4) & -40.0$^d$ (-3.6) \\
\hline
\multirow{3}{*}{\textbf{DSD-PBEP86-D3}$^e$}   & 2 & 2               & 68.1             & -51.5            \\
                                    & 3 & 3              & 48.3 (-19.8)     & -49.6  (-1.9).   \\ 
                                    & 4 & 4              & 46.1 (-22.0)     & -47.8  (-3.7)    \\
                                    & 20 & 4             & 32.3$^f$ (-35.8) & -43.7$^f$ (-7.8) \\
                                    &   &                  & 32.1$^g$ (-36.0) & -41.7$^g$ (-9.8) \\ 
                                    & 192 & 4          & 32.4$^h$ (-35.7) & -45.7$^h$ (-6.8) \\
\hline

\multirow{2}{*}{\textbf{jun-ChS}}  &  2 & 2            & 78.3         & -50.0        \\
                                    & 3 & 3           & 58.5 (-19.8)$^i$ & -48.0 (-2.0)$^i$ \\
                                    & 20 & 4         & 44.1 (-34.2)$^j$ & -40.8 (-9.2)$^j$ \\ 
\hline
\end{tabular}

$^a$ In parentheses is the difference with respect to \ce{(H2O)2} results.\\
$^b$ 6-31+G(d,p) basis set as in ref \cite{Koch-2007} . \\
$^c$ jul-cc-pVDZ basis set.\\
$^d$ QM/MM energies and ZPVEs. 20 waters molecules treated at PW6B95-D3 level, the remaining molecules described by the Amber force field.\\
$^e$ jul-cc-pVTZ basis set.\\ 
$^f$ ONIOM geometries and ZPVE. DSD-PBEP86/jul-cc-pVTZ for adsorbate and molecules involved in the relay mechanism, PW6B96-D3/jul-cc-pVDZ for the water molecules not involved in the relay mechanism.\\
$^g$ DSD-PBEP86/jul-cc-pVTZ energies while geometries and ZPVE at PW6B95-D3/jul-cc-pVDZ level. \\
$^h$ DSD-PBEP86:PW6B95-D3:Amber energies on PW6B95D3:Amber geometries. ZPEs at PW6B95D3:Amber level.\\
$^i$ jun-ChS electronic energy, PW6B95-D3/jul-cc-pVDZ geometry and ZPVE.\\
$^j$ jun-ChS:PW6B95 electronic energy, PW6B95-D3/jul-cc-pVDZ geometry and ZPVE.
\end{table}

The dependence of the energy profile on the number of water molecules involved in the relay mechanism was already pointed out\cite{Gardebien-2003,Koch-2007}. However, only few water molecules were considered and no attempt to simulate the effect of ice bulk has been reported beyond the PCM level, whose reliability is, however, questionable for hydrogen-bonding solids. For comparison, Table \ref{table:SI_formationenergies}

lists the relative energies (corrected for the zero point vibrational energies, ZPVEs of the elementary steps obtained by Koch et al.\cite{Koch-2007} and in the present work (further details are given in Table S4 of the SI). As it can be seen, the relative energy for HCN interacting with the water cluster is only marginally affected by the cluster size, but there is a huge effect on the activation barrier. While an overall fair agreement between the present results and those obtained in ref. \citep{Koch-2007} can be noted, there is a difference of about 18 kJ/mol for the energy of the transition state. This can be explained by considering that Koch et al.\cite{Koch-2007} investigated the role of the crystalline environment by optimizing for the different stationary points the positions of seven water molecules around the \ce{HNC\bond{...}(H2O)3} complex without any constraint related to the arrangement of water molecules in icy structures. The importance of the morphological pattern in ice is highlighted by the present results: indeed, using a \ce{(H2O)20} cluster with the same molecular arrangement as in ice XI rules out the possibility of a process catalysed by two or three water molecules. Rather, the molecular arrangement at the surface permits a process assisted by four water molecules (see Figure  \ref{fig:PES_H2O20-ONIOM}).

Further support to the reliability of the results is provided by the comparable barrier obtained by another ONIOM computation in which the high-level part of the system (\ce{HCN\bond{...}(H2O)4}) is treated at the jun-ChS instead of DSD-PBEP86-D3 level without any additional geometry optimization (last line of Table \ref{table:SI_formationenergies}). 
What is even more gratifying is that the differences between the results obtained for the smallest \ce{HCN\bond{...}(H2O)2} model and the larger model clusters (values in parenthesis in Table \ref{table:SI_formationenergies}) obtained at the DSD-PBEP86 level are in quantitative agreement with the jun-ChS counterparts. This paves the route toward the computation of very reliable parameters for reactions occurring on icy grains by combining jun-ChS results for small models and ONIOM(DSD-PBEP86:PW6B95-D3) values for large model clusters. 

\begin{figure}[H]

    \centering
    \includegraphics[width=14cm]{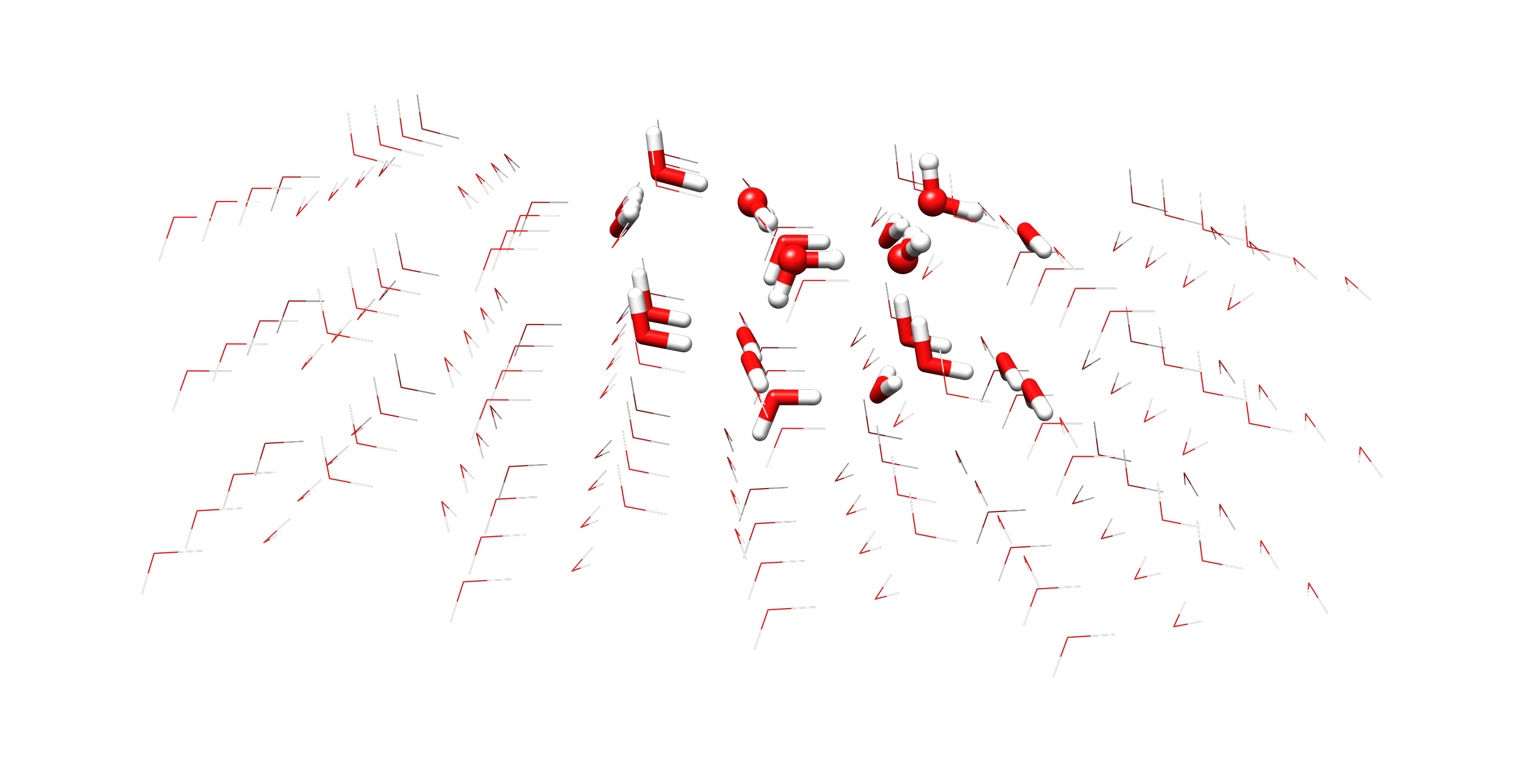}
    \caption{Structural model for the \ce{(H2O)192} cluster treated by three-layer ONIOM DSD-PBEP86:PW6B95:Amber strategy (geometry at PW6B95:Amber level). Ball and stick and tubular representation for the QM sections treated at DSD-PBEP86/jul-cc-pVTZ and PW6B95-D3/jul-cc-pVDZ level, respectively.}
    \label{fig:cluster192}

\end{figure}

This approach can be further extended to very large models by employing a three-layer QM/QM$'$/MM ONIOM model. In order to analyze also this aspect, we have embedded the \ce{HCN\bond{...}(H2O)20} cluster in a large model of ice-XI containing 172 water molecules described by the Amber force field (see Figure \ref{fig:cluster192}). The results collected in Table \ref{table:SI_formationenergies}
show that inclusion of the MM layer further stabilizes the HCN isomer with respect to the HNC counterpart by about 4 kJ/mol, but has a negligible effect on the energy barrier (less than 0.4 kJ/mol). Taking into account the estimated error bar of the overall computational approach (about 4 kJ/mol), the results obtained for the \ce{HCN\bond{...}(H2O)20} model can be considered essentially converged with respect to further extension of the ice substrate.

\subsection{Reaction rates}
In an astrochemical context HNC can either isomerize to HCN or diffuse on ice surfaces and then react with another molecule (e.g. CH$_2$NH to produce acetonitrile) at the low temperatures typical of the ISM. 

\begin{figure}[!ht]
    \centering
    \includegraphics[width=\textwidth]{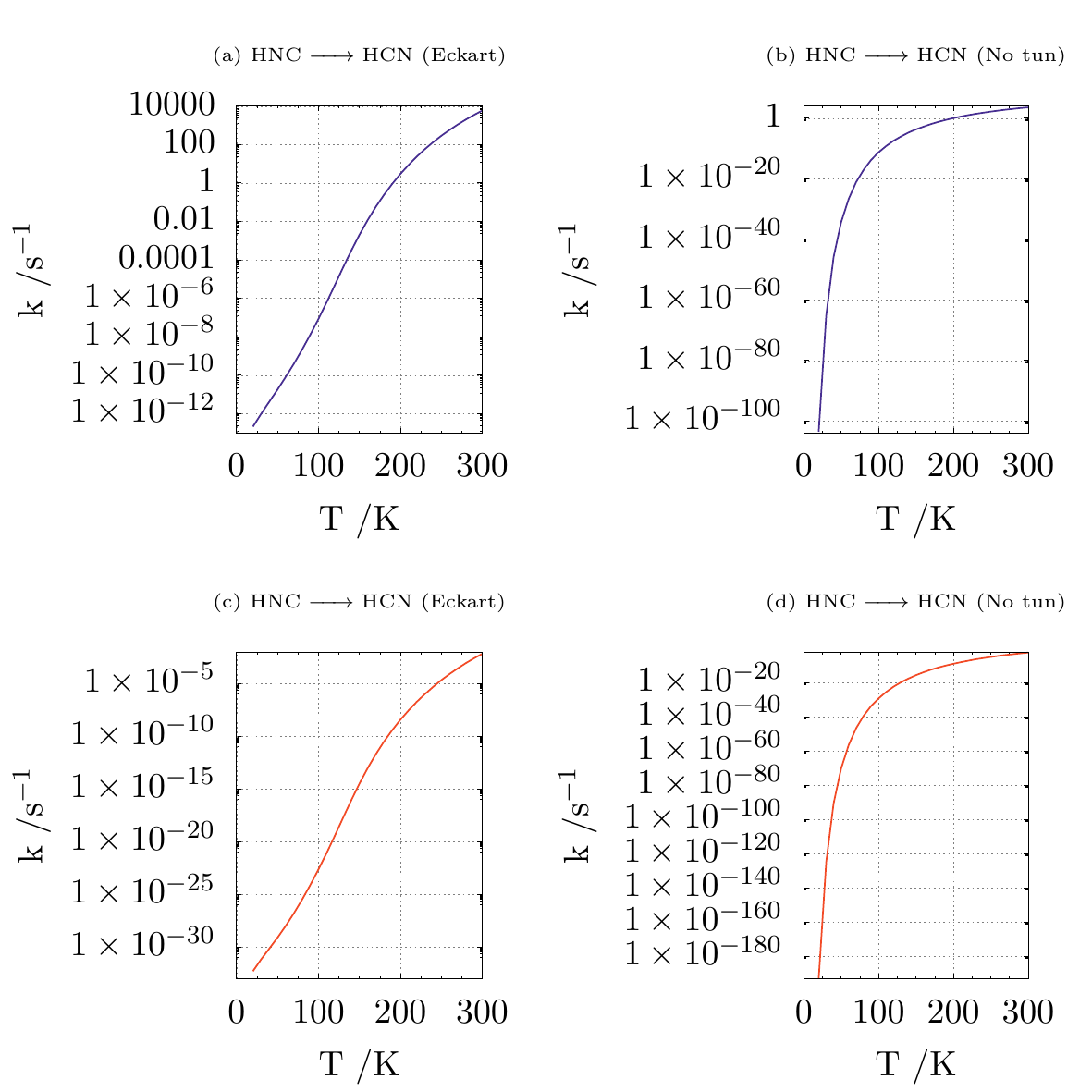}
    \caption{Reaction rates for the HNC$\rightleftharpoons$HCN isomerization including (Eckart) or excluding (No tun) tunneling. Panels a) and b) refer to the \ce{HNC\bond{...}(H2O)20} model, whereas panels c) and d) refer to the \ce{HNC\bond{...}(H2O)2} model.} 
    \label{fig:rate_tun_notun}
\end{figure}

The reaction rates computed for the HNC$\rightleftharpoons$HCN isomerization with the methodology described in Section 2 are shown in Figure \ref{fig:rate_tun_notun}. It is apparent that the rates computed for the \ce{HNC\bond{...}(H2O)2} system (Figure \ref{fig:rate_tun_notun}, panels c and d) are very slow irrespective of the inclusion or not of tunneling. The situation is completely different for the \ce{HNC\bond{...}(H2O)20} model, where the rate not including tunneling (corresponding to the one used by Koch and coworkers \cite{Koch-2007}) remains very small at low temperatures (Figure \ref{fig:rate_tun_notun}b), but inclusion of tunneling (Figure \ref{fig:rate_tun_notun}a) permits an effective reaction even at temperatures characteristic of the ISM. Noted is that the rates computed taking tunneling into account show a clear bimodal shape and cannot be fitted by a simple Arrhenius (or Kooij) function.\cite{laidler96,kooij}

Unfortunately, the diffusion coefficients of HNC (or even HCN) on ice have not yet been reported.\cite{hutwelker2006} According to a recent classification of ice adsorbates \cite{devlin1997}, HCN (hence probably HNC) is assigned to the intermediate class, which induces some deformation of the surface, but does not form hydrates nor penetrates rapidly into the ice bulk. An upper limit to the surface diffusion coefficient can be estimated with reference to the guess of \SI{4d-11}{\centi\metre\squared\per\second} at 130 K provided by Livingston et al. for \ce{SO2}, \cite{livingston2002} which corresponds to a mean distance of \SI{1260}{\angstrom } in \SI{1}{\second}.  Since the computed isomerization rate at 130 K is about 10$^{-4}$ s$^{-1}$ (which lowers to \SI{1d-10}{\per\second} at 50 K), the average diffusion of HNC before isomerization can reach \SI{100}{\angstrom} at 130 K (1 cm at 50 K).

 Therefore, if the formation of aminoacetonitrile is faster than the isomerization to HCN when HNC and \ce{CH2NH} are nearest neighbors,\cite{Koch-2008_Aminoacetonitrile} our results suggest that diffusion of HNC along significant distances could permit the formation of aminoacetonitrile on icy grains containing \ce{CH2NH} even at low concentrations. 
 
\subsection{Conclusions and outlook}

The main aim of this work was the implementation and validation of a general computational strategy for the study of the thermochemistry and kinetics of chemical processes taking place on interstellar icy-grains. To this end composite methods rooted in the coupled cluster ansatz have been combined with hybrid and double hybrid functionals together with molecular mechanics force field to characterize the stationary points ruling the reactive potential energy surfaces on model clusters sufficiently large to minimize spurious boundary effects. Next powerful master equation / RRKM models have been employed to compute reaction rates including tunneling effects. As a demanding test case we have selected the HCN/HNC reactions for which the available computational results are not fully satisfactory. \\
Ten different (meta-)hybrid and double-hybrid density functionals have been considered in conjunction with the jun-, jul- and aug-cc-pV$n$Z basis sets of double- and triple-$\zeta$ quality and their accuracy in predicting geometries together with thermochemical and kinetic 
data (adsorption, activation and reaction energies) has been assessed in comparison to reference values computed using the jun-ChS composite method. This benchmark has led to the conclusion that, among (meta-)hybrid functionals, BMK-D3 and PW6B95-D3 in conjunction with partially augmented double- and triple-$\zeta$ basis sets yield the most reliable description of geometries, with the optimal trade-off between accuracy and computational cost being offered by the PW6B95-D3/jul-cc-pVDZ model chemistry. Concerning double-hybrids, DSD-PBEP86-D3 and revDSD-PBEP86-D3 in conjunction with the jul-cc-pVTZ basis set deliver accurate predictions of both geometries and reaction energies. 
Next, these outcomes have been used to investigate the effect of cluster size and ice surface on the isomerization process of HCN. In particular, a cluster containing 20 water molecules has been cut from the (0 1 0) surface of ice XI, and used in a multiscale ONIOM calculation, in which the reaction center has been modeled at the DSD-PBEP86-D3/jul-cc-pVTZ level, while for the remaining portion of the \ce{(H2O)20} cluster the PW6B95-D3 functional has been employed in conjunction with the jul-cc-pVDZ basis set. This approach has allowed the proper modelling of the surface with an accurate yet cost-effective strategy. The pivotal role of the structural arrangement of surface molecules in driving the evolution of catalytic processes has been pointed out. The accuracy of the results has been further improved by combining jun-ChS results for small models to QM/QM' (DSD-PBEP86:PW6B95-D3) values for medium size model clusters and/or three-layer QM/QM$'$/MM computations for very large clusters. \\
On top of these computations, reaction rates have been computed by methods rooted in the transition state theory including tunneling which plays the dominant role at low temperature for processes involving the motion of light atoms. At variance with previous investigations, our results show that the isomerization is ruled by a proton relay mechanism directly involving four water molecules, but tuned by relatively distant waters belonging to the model cluster employed to mimic the ice surface. The resulting activation energy is strongly reduced with respect to that governing the isomerization of the bare HCN molecule, but only tunneling allows for effective isomerization of HNC in the harsh conditions characterizing astrochemical processes. \\
Together with the intrinsic interest of the studied system, the results of the present work have allowed to define the best strategy for future modelling of iCOMs-ices interactions in the framework of a QM/QM$'$/MM approach. This also represents the starting point for hybrid QM/QM$'$/periodic approaches, in which the outcome of the multiscale (QM/QM$'$) description of the cluster is corrected for environmental effects obtained by simulating the surface using periodic boundary conditions \cite{Sauer2019}. However, the crystalline water ice surfaces usually employed to simulate icy dust grains could be inadequate to describe their amorphous structure. Work in this and related directions is under way in our laboratory in order to achieve a more realistic modeling of chemical processes occurring on icy mantels of interstellar grains.

\begin{acknowledgement}
This work has been supported by MIUR (Grant Number 2017A4XRCA), by the Italian Space Agency (ASI; ‘Life in Space’ project, N. 2019-3-U.0) and by Scuola Normale Superiore (SNS18$_-$B$_-$Tasinato). The SMART@SNS Laboratory (\url{http://smart.sns.it}) is acknowledged for providing high-performance computing facilities.
\end{acknowledgement}

\begin{suppinfo}
Main structural parameters for the water dimer obtained at the ChS and jun-ChS levels. 
Error analysis for the structural parameters of the \ce{HCN\bond{...}(H2O)2} system.
Contributions to jun-ChS electronic energies. Error analysis for the formation energy of the species involved in HNC$\rightleftharpoons$HCN isomerization assisted by the \ce{(H2O)2} cluster. Relative ground state electronic energies for the stationary points on the HCN$\rightleftharpoons$HNC isomerization PES with respect to isolated HCN and (H$_2$O)$_n$ for $n = 2, 20$.
Cartesian coordinates of the stationary points on the HCN$\rightleftharpoons$HNC isomerization PES on a \ce{(H2O)}$_n$ cluster for $n$ = 2, 3 and 4 optimized at the DSDPBEP86-D3/jul-cc-pVTZ level of theory. Cartesian coordinates of the stationary points on HCN$\rightleftharpoons$HNC isomerization PES catalyzed by the \ce{(H2O)20} cluster optimized at the DSDPBEP86-D3/jul-cc-pVTZ:PW6B95-D3/jul-cc-pVDZ level of theory
This material is available free of charge via the Internet at http://pubs.acs.org.
\end{suppinfo}
\pagebreak

\bibliography{MyBib.bib}

\end{document}


\newpage
\section{List of supporting information}
\begin{figure}[h]
    \centering
    \includegraphics[width=8cm]{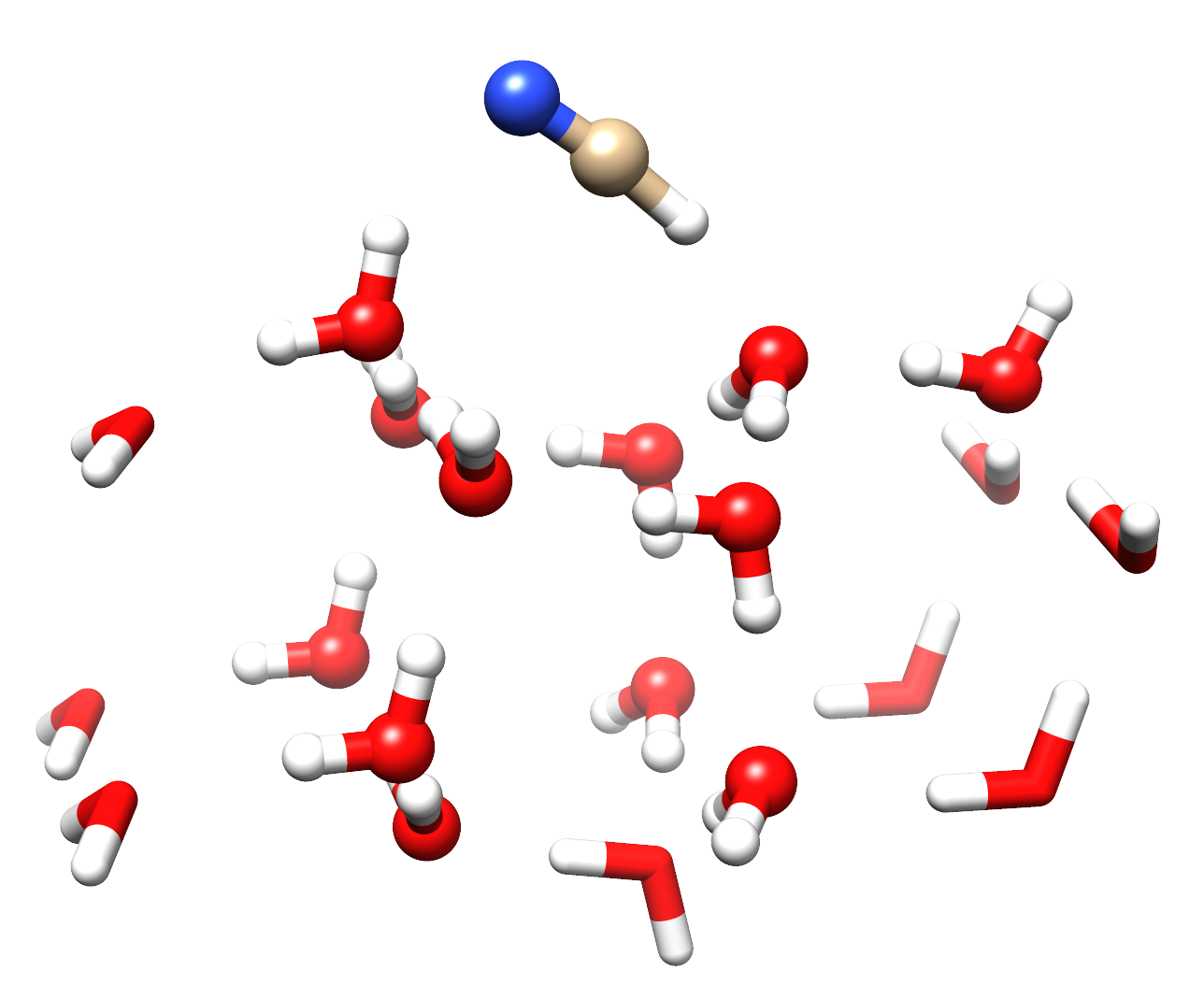}
    \caption{Structure of the HCN@\ce{(H2O)} system showing molecules frozen during the optimization. Ball and stick representation used for atoms free to move while tubular representation for molecules kept frozen in order to prevent structural distortion of the cluster.}
    \label{fig:SI_frozen_mol}
\end{figure}

\begin{itemize}
    \item [] Table \ref{table:SI_water}. Main structural parameters for the water dimer obtained at the ChS and jun-ChS levels and comparison to CCSD(T)-F12b/CBS+fT+fQ+CV+REL+DBOC values
    \item [] Table \ref{table:SI_MAEandRE}. Mean Absolute Errors and Relative Errors for bond lengths, angles and dihedrals of the \ce{HCN\bond{...}(H2O)2} system.
    \item [] Figure \ref{fig:MAEandRE}. Mean Absolute Errors (MAE) and Relative Errors (RE) for bond lengths, valence and dihedral angles.
    \item [] Table \ref{table:jchscontributions}. Contributions to jun-ChS electronic energies.
    \item [] Table \ref{table:SI_MAE-FormationEnergy}. Absolute Errors and Mean Absolute Errors of formation energies computed at the jun-ChS level on top of different optimized geometries with respect to full jun-ChS results.
    \item [] Figure \ref{fig:SI_frozen_mol}. Structure of the HCN@\ce{(H2O)} system showing molecules frozen during the optimization.
    \item [] Table \ref{tab:formation_energy_reac}. Relative ground state electronic energies for the stationary points on the HCN$\rightleftharpoons$HNC isomerization PES with respect to isolated HCN and (H$_2$O)$_n$ for $n = 2, 20$.
    \item [] Table \ref{table:SI_cartesian}. Cartesian coordinates of the stationary points on the HCN$\rightleftharpoons$HNC isomerization PES on a \ce{(H2O)}$_n$ cluster for $n$ = 2, 3, 4 optimized at the DSDPBEP86-D3/jul-cc-pVTZ level of theory.
    \item [] Table \ref{table:SI_cartesian}. Cartesian coordinates of the stationary points on HCN$\rightleftharpoons$HNC isomerization PES catalyzed by the \ce{(H2O)20} cluster optimized at the DSDPBEP86-D3/jul-cc-pVTZ:PW6B95-D3/jul-cc-pVDZ level of theory.

\end{itemize}

\newpage
\begin{table}[H]
\caption{Main structural parameters for the water dimer obtained at the ChS and jun-ChS levels and comparison to CCSD(T)-F12b/CBS+fT+fQ+CV+REL+DBOC values of ref.\cite{Lane-2013-bis} $\theta$$_a$ and $\theta$$_d$ are the H-O-H valence angles for the acceptor and the donor respectively, $\alpha$ gives a measure of the deviation from a linear hydrogen bond and $\beta$ gives the orientation of the $C_2$ axis of the acceptor with respect to the O-O axis. Bond lengths in Å and angles in degrees.}
\label{table:SI_water}
\begin{tabular}{lccc}
\hline
       & \textbf{ChS} & \textbf{jun-ChS} & \textbf{Ref. value}\cite{Lane-2013-bis} \\
\hline
\textbf{r(O-O)} & 2.9049 & 2.9058 & 2.9092       \\
\boldmath{$\theta$}$_{a}$   &104.78 & 104.77 & 104.95       \\
\boldmath{$\theta$}$_{d}$  &104.87  & 104.84 & 104.85       \\
\boldmath{$\alpha$}  & 4.91 & 5.91   & 5.69         \\
\boldmath{$\beta$}    & 126  & 127     & 123.46      \\
\hline
\end{tabular}
\end{table}

\pagebreak

\begin{longtable}{p{3.8cm}p{1.8cm}ccc|cccc}
\caption{MAE and RE for bond lengths (\textbf{r}), angles (\boldsymbol{$\alpha$}) and dihedrals (\boldsymbol{$\upphi$}) of the 
HCN--\ce{(H2O)2} system. Averaged values over all the species along the PES are collected. Last column collect total RE obtained averaging over all the structural parameters. MAE for bond lengths in Å and in degrees for angles and dihedrals. RE are in \%.}

\label{table:SI_MAEandRE}\\
\hline
\multicolumn{2}{l}{\textbf{}} & \multicolumn{3}{c|}{\textbf{MAE}} & \multicolumn{4}{c}{\textbf{RE}} \\
\hline
\multicolumn{2}{c}{\textbf{}} & \textbf{r} & \boldsymbol{$\alpha$} & \boldsymbol{$\upphi$} & \textbf{r} & \boldsymbol{$\alpha$} & \boldsymbol{$\upphi$} & \textbf{Total}\\ 
\hline
\multirow{6}{*}{\textbf{B3LYP-D3}}       & \textbf{Jun-DZ} & 0.028 & 1.72 & 2.21 & 1.78 & 1.31 & 2.43  & 1.02\\
                                         & \textbf{Jul-DZ} & 0.023 & 1.45 & 3.56 & 1.47 & 1.23 & 3.65  & 0.98\\
                                         & \textbf{Aug-DZ} & 0.021 & 1.24 & 2.53 & 1.36 & 1.08 & 2.61  & 0.88 \\
                                         & \textbf{Jun-TZ} & 0.013 & 1.07 & 2.69 & 0.87 & 0.89 & 3.08  & 0.69\\
                                         & \textbf{Jul-TZ} & 0.013 & 1.03 & 2.44 & 0.86 & 0.86 & 2.69  & 0.65\\
                                         & \textbf{Aug-TZ} & 0.013 & 1.09 & 2.51 & 0.89 & 0.92 & 2.82  & 0.69\\
\hline
\multirow{6}{*}{\textbf{BHLYP-D3}}       & \textbf{Jun-DZ} & 0.092 & 7.34 & 9.21 & 3.50 & 6.17 & 10.47 &4.82\\
                                         & \textbf{Jul-DZ} & 0.016 & 1.64 & 2.52 & 0.87 & 1.39 & 2.38  &0.85\\
                                         & \textbf{Aug-DZ} & 0.014 & 1.46 & 1.64 & 0.79 & 1.25 & 1.81  &0.76\\
                                         & \textbf{Jun-TZ} & 0.014 & 1.28 & 1.59 & 0.96 & 1.07 & 1.65  & 0.80\\
                                         & \textbf{Jul-TZ} & 0.014 & 1.34 & 1.54 & 0.95 & 1.10 & 1.82  & 0.83\\
                                         & \textbf{Aug-TZ} & 0.015 & 1.34 & 1.48 & 0.97 & 1.11 & 1.72  & 0.83\\
\hline
\multirow{6}{*}{$\mathbf{\omega}$\textbf{B97X-D}}       & \textbf{Jun-DZ} & 0.030 & 2.36 & 4.01 & 1.65 & 1.90 & 3.28  &1.06\\
                                         & \textbf{Jul-DZ} & 0.023 & 1.82 & 5.01 & 1.25 & 1.57 & 4.86 & 0.91\\
                                         & \textbf{Aug-DZ} & 0.021 & 1.70 & 5.15 & 1.14 & 1.48 & 4.89  &0.92\\
                                         & \textbf{Jun-TZ} & 0.017 & 1.63 & 5.06 & 0.88 & 1.37 & 5.23  &0.81\\
                                         & \textbf{Jul-TZ} & 0.017 & 1.59 & 4.87 & 0.88 & 1.34 & 4.95  &0.78\\
                                         & \textbf{Aug-TZ} & 0.018 & 1.62 & 5.00 & 0.93 & 1.38 & 5.17  &0.81\\
\hline
\multirow{6}{*}{\textbf{PW6B95-D3}}      & \textbf{Jun-DZ} & 0.022 & 2.03 & 2.47 & 1.23 & 1.50 & 2.48 & 1.00\\
                                         & \textbf{Jul-DZ} & 0.015 & 1.34 & 3.11 & 0.80 & 1.09 & 3.05 & 0.74\\
                                         & \textbf{Aug-DZ} & 0.013 & 1.05 & 1.96 & 0.73 & 0.90 & 1.65 & 0.60\\
                                         & \textbf{Jun-TZ} & 0.011 & 0.96 & 2.56 & 0.55 & 0.79 & 2.59 & 0.55\\
                                         & \textbf{Jul-TZ} & 0.010 & 0.94 & 2.31 & 0.53 & 0.76 & 2.29 & 0.52\\
                                         & \textbf{Aug-TZ} & 0.010 & 0.95 & 2.34 & 0.52 & 0.78 & 2.36 & 0.53\\
\hline
\multirow{6}{*}{\textbf{BMK-D3}}         & \textbf{Jun-DZ} & 0.015 & 1.87 & 3.32 & 1.00 & 1.36 & 2.71 & 0.75\\
                                         & \textbf{Jul-DZ} & 0.013 & 1.67 & 4.09 & 0.84 & 1.38 & 3.84 & 0.77\\
                                         & \textbf{Aug-DZ} & 0.012 & 1.60 & 2.84 & 0.77 & 1.34 & 2.47 & 0.69\\
                                         & \textbf{Jun-TZ} & 0.008 & 1.03 & 2.32 & 0.52 & 0.84 & 2.04 & 0.57\\
                                         & \textbf{Jul-TZ} & 0.008 & 1.00 & 2.00 & 0.52 & 0.81 & 1.61 & 0.54\\
                                         & \textbf{Aug-TZ} & 0.008 & 1.03 & 2.02 & 0.51 & 0.84 & 1.68 & 0.56\\
\hline
\multirow{6}{*}{\textbf{M06-2X}}         & \textbf{Jun-DZ} & 0.014 & 1.95 & 2.65 & 1.05 & 1.37 & 2.99 & 1.09\\
                                         & \textbf{Jul-DZ} & 0.013 & 1.16 & 2.96 & 0.91 & 0.96 & 2.72 &  0.85\\
                                         & \textbf{Aug-DZ} & 0.014 & 1.18 & 2.62 & 0.90 & 0.96 & 2.74 & 0.84\\
                                         & \textbf{Jun-TZ} & 0.012 & 1.05 & 2.36 & 0.72 & 0.82 & 2.06 & 0.70\\
                                         & \textbf{Jul-TZ} & 0.012 & 1.00 & 1.90 & 0.74 & 0.79 & 1.39 & 0.65\\
                                         & \textbf{Aug-TZ} & 0.012 & 1.04 & 1.86 & 0.75 & 0.83 & 1.41&  0.66\\
\hline
\multirow{6}{*}{\textbf{MN15}}           & \textbf{Jun-DZ} & 0.017 & 2.15 & 3.08 & 1.18 & 1.59 & 2.62 & 1.15\\
                                         & \textbf{Jul-DZ} & 0.013 & 1.49 & 3.70 & 0.94 & 1.17 & 3.23  &0.95\\
                                         & \textbf{Aug-DZ} & 0.010 & 1.20 & 2.80 & 0.82 & 0.99 & 2.87 & 0.83\\
                                         & \textbf{Jun-TZ} & 0.009 & 1.03 & 3.13 & 0.63 & 0.82 & 2.48 & 0.68\\
                                         & \textbf{Jul-TZ} & 0.008 & 1.04 & 2.69 & 0.60 & 0.82 & 1.99 & 0.65\\
                                         & \textbf{Aug-TZ} & 0.008 & 1.01 & 2.60 & 0.58 & 0.80 & 1.93 & 0.65\\
\hline
\multirow{6}{*}{\textbf{B2PLYP-D3}}      & \textbf{Jun-DZ} & 0.025 & 1.88 & 2.36 & 1.58 & 1.39 & 2.89 & 1.10\\
                                         & \textbf{Jul-DZ} & 0.020 & 1.23 & 2.71 & 1.29 & 1.03 & 2.62 & 0.88\\
                                         & \textbf{Aug-DZ} & 0.018 & 1.00 & 1.74 & 1.21 & 0.88 & 1.65 & 0.75\\
                                         & \textbf{Jun-TZ} & 0.010 & 0.77 & 1.65 & 0.63 & 0.63 & 1.77 & 0.51\\
                                         & \textbf{Jul-TZ} & 0.009 & 0.69 & 2.13 & 0.58 & 0.57 & 2.72 & 0.59\\
                                         & \textbf{Aug-TZ} & 0.009 & 0.74 & 1.50 & 0.60 & 0.63 & 1.60&  0.48\\
\hline
\multirow{6}{*}{\textbf{DSD-PBEP86-D3}}  & \textbf{Jun-DZ} & 0.021 & 1.78 & 2.48 & 1.44 & 1.28 & 3.02  & 1.07\\
                                         & \textbf{Jul-DZ} & 0.017 & 0.88 & 2.63 & 1.18 & 0.74 & 2.60& 0.76 \\
                                         & \textbf{Aug-DZ} & 0.016 & 0.96 & 2.39 & 1.15 & 0.84 & 2.64 & 0.81\\
                                         & \textbf{Jun-TZ} & 0.008 & 0.54 & 1.50 & 0.60 & 0.44 & 1.69 & 0.43\\
                                         & \textbf{Jul-TZ} & 0.007 & 0.46 & 1.39 & 0.57 & 0.37 & 1.55 & 0.41\\
                                         & \textbf{Aug-TZ} & 0.007 & 0.59 & 1.53 & 0.58 & 0.49 & 1.73 & 0.44\\
\hline
\multirow{6}{*}{\textbf{rDSD-PBEP86-D3}} & \textbf{Jun-DZ} & 0.023 & 1.80 & 2.53 & 1.49 & 1.31 & 2.98 & 1.11\\
                                         & \textbf{Jul-DZ} & 0.017 & 0.89 & 2.64 & 1.15 & 0.75 & 2.58&  0.77\\
                                         & \textbf{Aug-DZ} & 0.016 & 0.96 & 2.38 & 1.12 & 0.83 & 2.64&  0.80\\
                                         & \textbf{Jun-TZ} & 0.008 & 0.52 & 1.51 & 0.58 & 0.42 & 1.66 & 0.43\\
                                         & \textbf{Jul-TZ} & 0.007 & 0.43 & 1.36 & 0.54 & 0.35 & 1.50 & 0.40\\
                                         & \textbf{Aug-TZ} & 0.007 & 0.54 & 1.53 & 0.55 & 0.45 & 1.73 & 0.42\\
                                         
\hline
\end{longtable}
\pagebreak

\begin{figure}[h!]
    \centering
    \includegraphics[width=\textwidth]{Figures/MAE&RE-distance&angle&dihedrals-SumOverSpecies.png}
    \caption{Mean Absolute Errors (MAE) and Relative Errors (RE) for bond lengths, valence and dihedral angles. The values are obtained by averaging absolute and relative errors of structural parameters over all the species involved in the reactive PES for the HCN$\rightleftharpoons$HNC isomerization assisted by two water molecules.}
    \label{fig:MAEandRE}

\end{figure}
\pagebreak

{\renewcommand{\arraystretch}{1.8}
\begin{table}[]
\resizebox{\textwidth}{!}
    {\begin{tabular}{p{5.8cm}lcccccccc}
    \hline
            &      \textbf{HCN} & \textbf{\ce{(H2O)$_2$}} & \textbf{HCN--\ce{(H2O)$_2$}} & \textbf{TS}   & \textbf{HNC--\ce{(H2O)$_2$}} & \textbf{HNC}\\
        \hline
        \textbf{E(CC)}& -93.2774 & -152.6856 & -245.9753 & -245.9254 & -245.9574 & -93.2540 \\
        \textbf{$\Delta$E$_\textbf{MP2}^\textbf{$\infty$}$}
         & -0.0446 & -0.0818 & -0.1268 & -0.1262 & -0.1264 & -0.0444 \\
        \textbf{$\Delta$E$_\textbf{MP2}^\textbf{CV}$} & -0.0944 & -0.1029 & -0.1974 & -0.1974 & -0.1973 & -0.0943 \\
        \hline
        \textbf{E(jun-ChS)} & -93.4165 & -152.8703 & -246.2995 & -246.2490 & -246.2810 & -93.3927 \\
    
        \hline
    \end{tabular}} \quad
    \caption{jun-ChS contributions (in Hartree) to energy evaluated on jun-ChS geometries. E(CC) is the CCSD(T) energy computed with the jun-cc-pVTZ basis set; $\Delta$E$_{MP2}^{\infty}$ is the difference between the fc-MP2/jun-cc-pVTZ energy and the corresponding extrapolated value estimated by the jun-cc-pVnZ basis sets with n = T and Q. $\Delta$E$_{MP2}^{CV}$ accounts for core-valence correlation and is obtained as difference between ae- and fc- MP2 calculations with cc-pwCVTZ basis set. The final jun-ChS energy is reported in the last row.}
    \label{table:jchscontributions}
\end{table}}

\begin{longtable}{p{5.8cm}cccc|c}

\caption{AE and MAE of formation energies computed at the jun-ChS level on top of different optimized geometries with respect to full jun-ChS results. The formation energy of each species (in kJ/mol) is calculated with respect to isolated HCN + \ce{(H2O)2}. RC stands for reactant complex, i. e. NCH--\ce{(H2O)2}, TS for transition state and PC for product complex, i. e. CNH--\ce{(H2O)2}.}
\label{table:SI_MAE-FormationEnergy} \\
\hline
 & \textbf{RC} & \textbf{TS} & \textbf{PC} & \textbf{HCN + \ce{(H2O)2}} & \textbf{MAE} \\
\hline
\textbf{PW6B95-D3/jul-DZ}      & 0.04       & 0.11     & 0.20  &0.04  &0.10\\
\hline
\textbf{BHLYP-D3/aug-DZ}       & 0.01                          & 0.05                   & 0.16                          & 0.39                           & 0.15           \\
\textbf{PW6B95-D3/aug-DZ}      & 0.02                          & 0.15                   & 0.13                          & 0.01                           & 0.08           \\
\textbf{BMK-D3/aug-DZ}         & 0.29                          & 0.12                   & 0.00                          & 0.01                           & 0.11           \\
\textbf{M06-2X/aug-DZ}         & 0.09                          & 0.41                   & 0.34                          & 0.01                           & 0.21           \\
\textbf{MN15/aug-DZ}           & 0.03                          & 0.11                   & 0.57                          & 0.36                           & 0.27           \\
\hline
\textbf{PW6B95-D3/jul-TZ}      & 0.07                          & 0.11                   & 0.30                          & 0.38                           & 0.22           \\
\textbf{BMK-D3/jul-TZ}         & 0.05                          & 0.08                   & 0.45                          & 0.61                           & 0.30           \\
\textbf{M06-2X/jul-TZ}         & 0.04                          & 0.23                   & 0.12                          & 0.42                           & 0.20           \\
\textbf{MN15/jul-TZ}           & 0.09                          & 0.07                   & 0.05                          & 0.24                           & 0.11           \\
\textbf{DSD-PBEP86-D3/jul-TZ}  & 0.10                          & 0.09                   & 0.03                          & 0.05                           & 0.07           \\
\textbf{rDSD-PBEP86-D3/jul-TZ} & 0.12                          & 0.06                   & 0.02                          & 0.05                           & 0.06    \\
\hline
\end{longtable}

\pagebreak

\pagebreak

\begin{table}[]
\resizebox{\textwidth}{!}
    {\begin{tabular}{p{5.8cm}lccccc}
    \hline
                                        &      & \textbf{HCN--\ce{(H2O)$_n$}} & \textbf{TS}   & \textbf{HNC--\ce{(H2O)$_n$}} & \textbf{HNC + \ce{(H2O)}$_n$}\\
    \hline
        \multirow{6}{*}{\textbf{jun-ChS}}  & n=2 [2] & -33.4 & 99.3  & 15.0 & 62.3 \\
                                        & & \textbf{6.3} & \textbf{1.9} & \textbf{7.8} & \textbf{-1.3} \\
                                        & & (-27.2) & (101.2) & (22.8) & (61.0) \\
                                        & n = 4 [20] & -61.4 & 43.4 & -18.5 & 62.3\\
                                        & & \textbf{10.6} & \textbf{-9.3} & \textbf{8.6} & \textbf{-2.0} \\
                                        & & (-50.7) &  (34.2) &  (-9.9) & (60.2) \\
    \hline
                                        
        \multirow{3}{*}{\textbf{PW6B95}}& n=4 [20] & -64.1 & 27.6 & -27.8 & 56.7  \\
        & & \textbf{10.6} & \textbf{-9.3} & \textbf{8.6} & \textbf{-2.0} \\
        & & (-53.5) & (18.4) & (-19.2) & (54.7) \\
    \hline
        \multirow{3}{*}{\textbf{DSD-PBEP86}}  & n=4 [20] & -65.4 & 28.3 & -21.6 & 64.9 \\
        & & \textbf{7.9} & \textbf{-8.9} & \textbf{8.1} & \textbf{-1.3} \\
        & & (-57.5) & (19.4) &  (-13.4) & (63.6) \\
       \hline     
    \end{tabular}}
    \caption{Relative ground state electronic energies for the stationary points on the HCN$\rightleftharpoons$HNC isomerization PES with respect to isolated HCN and (H$_2$O)$_n$. The number of water molecules involved in the relay mechanism is explicitly indicated together with the total number of water molecules included in the model system (in parenthesis).  ZPVEs are in bold while energy corrected for ZPVEs are in parenthesis. Values in kJ/mol. jun-ChS energy and geometries and ZPVEs at DSD-PBEP86-D3/jul-cc-pVTZ for n=2. jun-ChS energies for the adsorbate and the water molecules involved in the proton relay and PW6B95D3/jul-cc-pVDZ for the remaining water molecules in the cluster for n=4 [20]. PW6B95-D3/jul-cc-pVDZ for PW6B95 results.
    DSD-PBEP86 results refers to ONIOM(DSD-PBEP86/jul-cc-pVTZ:PW6B95D3/jul-cc-pVDZ), both geometries and ZPVEs at this level.}
    \label{tab:formation_energy_reac}

\end{table}

\begin{longtable}{c}
\caption{Cartesian coordinates of the stationary points on the HCN$\rightleftharpoons$HNC isomerization PES on a \ce{(H2O)}$_n$ cluster for $n$ = 2, 3, 4 optimized at the DSDPBEP86-D3/jul-cc-pVTZ level of theory.} \\
\label{table:SI_cartesian}\\
\hline
\textbf{\ce{HCN\bond{...}(H2O)_2}} \\
\hline
H -2.4172460000 -1.3236170000 -0.3753170000 \\
H -1.4926540000 -0.1512860000 0.0046880000 \\
O -1.5878820000 -1.1174680000 0.0614470000  \\
H 0.2173170000 1.4761330000 -0.1085270000  \\
H -0.8952870000 2.2488950000 0.6011570000  \\
O -0.7411690000 1.5897050000 -0.0809160000  \\
C 1.3176910000 -0.7998110000 0.0131070000  \\
H 0.5103540000 -1.5088020000 0.0288850000  \\
N 2.1148270000 0.0399500000 -0.0105400000  \\
\hline
\textbf{TS@\ce{(H2O)_2}} \\
\hline
C -1.2945800000 -0.8862040000 0.0452440000 \\
H 0.3338370000 -1.2836170000 0.1070700000 \\
N -1.6744600000 0.2242890000 -0.0376320000 \\
H -0.3449010000 1.1193430000 -0.1006790000 \\
O 1.3191180000 -0.9595600000 0.0912560000 \\
O 0.6612240000 1.3491600000 -0.0925100000 \\
H 0.8305300000 1.8625930000 0.7062840000 \\
H 1.1046330000 0.2572530000 -0.0053710000 \\
H 1.7218600000 -1.3251680000 -0.7053130000 \\
\hline
\textbf{\ce{HNC\bond{...}(H2O)_2}} \\
\hline
C -2.2892940000 0.4017320000 -0.0124330000  \\
H 0.3174650000 1.5296530000 -0.0990770000  \\
N -1.3931210000 -0.3518240000 0.0118970000  \\
H -0.5499480000 -0.9398710000 0.0254200000  \\
O 1.2690700000 1.3714520000 -0.0833590000  \\
O 1.1166430000 -1.3928350000 0.0944840000  \\
H 1.5599600000 -1.8957890000 -0.5933260000  \\
H 1.4530840000 -0.4808450000 0.0269700000  \\
H 1.6213520000 2.0102850000 0.5423370000 \\
\hline
\textbf{\ce{HCN\bond{...}(H2O)_3}} \\
\hline
O -2.3584400000 0.5238630000 -0.0460450000 \\
H -1.8944140000 -0.3373700000 -0.0302560000 \\
H -3.0921230000 0.4155400000 -0.6550970000 \\
O -0.7934060000 -1.7514890000 0.0645970000  \\
H -0.8862590000 -2.3150170000 0.8367420000  \\
H 0.1662870000 -1.5849630000 -0.0138990000 \\
O 1.9246420000 -1.1641700000 -0.0551470000 \\
H 2.4942560000 -1.4374660000 -0.7788180000 \\
H 2.0268370000 -0.2025640000 0.0038430000  \\
C 0.2842930000 1.7579010000 0.0507490000  \\
H -0.7931660000 1.6577270000 0.0011210000  \\
N 1.4414940000 1.7701540000 0.0892320000 \\
\hline
\textbf{TS@\ce{(H2O)_3}} \\
\hline
O -2.1012760000 -0.2913010000 0.0379170000 \\
H -1.2555920000 0.7956740000 0.0165190000 \\
H -2.6039900000 -0.4087480000 0.8500230000 \\
O -0.5665470000 1.6243910000 -0.0630800000 \\
H -0.7096130000 2.0504520000 -0.9152650000 \\
H 0.5149180000 1.3128350000 0.0146840000 \\
O 1.7561250000 0.9443090000 0.0489610000 \\
H 2.1998450000 1.1557290000 0.8764210000 \\
H 1.7168340000 -0.0787420000 -0.0125640000 \\
C 0.0653410000 -1.8276070000 -0.0776250000 \\
H -1.4372350000 -1.0764350000 -0.0136650000 \\
N 1.2109090000 -1.5720440000 -0.0772550000 \\
\hline
\textbf{\ce{HNC\bond{...}(H2O)_3}} \\
\hline
O -2.1812290000 0.6508380000 0.0210110000 \\
H -0.5818510000 1.4812480000 0.0188820000 \\
H -2.8082780000 0.7829560000 0.7371090000 \\
O 0.3017000000 1.8990520000 -0.0250100000 \\
H 0.2527440000 2.5149750000 -0.7603410000 \\
H 1.6418500000 0.7824770000 0.0090730000 \\
O 2.2587060000 0.0165140000 0.0087320000 \\
H 2.8558970000 0.1515870000 0.7485510000 \\
H 1.1438380000 -1.2122110000 -0.0030890000 \\
C -0.8619140000 -2.1038320000 -0.0710360000 \\
H -2.0754540000 -0.3100620000 -0.0492460000 \\
N 0.2441880000 -1.7284580000 -0.0446550000 \\
\hline
\textbf{\ce{HCN\bond{...}(H2O)_4}} \\
\hline
O -2.8436820000 -0.7440310000 -0.0389370000 \\
H -3.5830050000 -0.6505780000 -0.6435400000 \\
H -2.4805420000 0.1586880000 0.0816930000 \\
H 1.1958030000 2.1797050000 -1.2964940000 \\
H 1.7053460000 1.2727840000 -0.1574010000 \\
O 1.0214320000 1.9333040000 -0.3852770000 \\
H -0.6675230000 1.7396320000 0.0638830000 \\
O -1.6156500000 1.6702220000 0.3027840000 \\
H -1.7042870000 2.1484650000 1.1303760000 \\
H 2.3982480000 -0.8863900000 0.2053580000 \\
O 2.8361970000 -0.0261130000 0.3163660000 \\
H 3.7516140000 -0.1581190000 0.0597530000 \\
C -0.1950600000 -1.9590570000 -0.0909090000 \\
H -1.2573540000 -1.7304940000 -0.1327920000 \\
N 0.9465260000 -2.1409150000 -0.0464100000 \\
\hline
\textbf{TS@\ce{(H2O)_4}} \\
\hline
O 1.8176240000 -0.7589480000 -1.2502330000 \\
H 2.7630220000 -0.9296150000 -1.3078420000 \\
H 1.0464500000 -1.2976530000 0.0229690000 \\
H -2.2791540000 -1.8623350000 -0.3765390000 \\
H -2.0530740000 -0.3134370000 -0.1386080000 \\
O -1.9098530000 -1.2269600000 0.2418240000 \\
H -0.5166790000 -1.3346420000 0.6023600000 \\
O 0.4968710000 -1.3021280000 0.8765870000 \\
H 0.6638200000 -0.3725560000 1.1918100000 \\
H -1.0244670000 1.4993950000 -0.2833040000 \\
O -1.9563890000 1.2618790000 -0.5212820000 \\
H -2.1235360000 1.6483900000 -1.3837640000 \\
C 1.6206590000 1.4263090000 0.9577060000 \\
H 1.7402660000 0.1775530000 -0.9798640000 \\
N 0.6390540000 1.4909000000 0.3044830000 \\
\hline
\textbf{\ce{HNC\bond{...}(H2O)_4}} \\
\hline
O -2.9134910000 -0.2603150000 -0.3272480000 \\
H -3.7881330000 -0.4487370000 0.0212990000 \\
H -1.8360840000 1.0970350000 0.1226290000 \\
H 1.5413260000 2.3638860000 -1.0133720000 \\
H 2.3984320000 0.3300010000 -0.0403770000 \\
O 1.4584710000 1.7588820000 -0.2731510000 \\
H 0.4988250000 1.7298220000 -0.0664370000 \\
O -1.1933760000 1.7966160000 0.3571760000 \\
H -1.3536570000 1.9869640000 1.2845570000 \\
H 1.4285180000 -1.4920410000 0.1149690000 \\
O 2.7653720000 -0.5620600000 0.1571390000 \\
H 3.4949620000 -0.6988150000 -0.4515920000 \\
C -0.6853020000 -2.1002670000 0.0419490000 \\
H -2.3963510000 -1.0790810000 -0.2119390000 \\
N 0.4554520000 -1.8646310000 0.0967480000 \\
\hline
\end{longtable}

\begin{longtable}{c}
\caption{Cartesian coordinates of the stationary points on HCN$\rightleftharpoons$HNC isomerization PES catalyzed by the \ce{(H2O)20} cluster optimized at the DSDPBEP86-D3/jul-cc-pVTZ:PW6B95-D3/jul-cc-pVDZ level of theory} \\
\label{table:SI_cartesian}\\
\hline
\textbf{\ce{HCN\bond{...}(H2O)_{20}}} \\
\hline
H 5.0223470000 0.6152780000 1.3926780000 \\
H 5.0186410000 0.5991620000 3.0286570000 \\
O 5.2779240000 0.0944080000 2.2061930000 \\
H -1.9111380000 1.9471350000 1.6036090000 \\
H -1.8348980000 2.0006570000 3.1507030000 \\
O -1.7156920000 1.4160370000 2.3996800000 \\
H 0.1653830000 0.9696300000 2.2828750000 \\
O 1.1094180000 0.8078430000 2.1577780000 \\
H 1.1991970000 -0.1621710000 2.1672440000 \\
H 4.5199270000 -1.3906580000 2.1901060000 \\
O 4.1402550000 -2.3164700000 2.1799070000 \\
H 3.1352740000 -2.2918190000 2.1780850000 \\
O 1.1494160000 -1.9732040000 2.2230220000 \\
H 0.6759950000 -2.3897110000 2.9450200000 \\
H 0.9487140000 -2.5015840000 1.4261470000 \\
H -2.3589060000 -0.3750370000 2.3360530000 \\
O -2.6377900000 -1.2978890000 2.2442720000 \\
H -3.6008880000 -1.3032660000 2.2955080000 \\
O -5.5380320000 -1.5969230000 2.1649560000 \\
H -5.9101890000 -2.0423080000 2.9837540000 \\
H -5.9064200000 -2.0262330000 1.3359450000 \\
H 5.0324660000 0.6580680000 -3.0031550000 \\
H 5.0286490000 0.6419680000 -1.3671390000 \\
O 5.2879850000 0.1373770000 -2.1895510000 \\
H 5.1655410000 2.5430170000 0.0313410000 \\
O 4.6629020000 1.7269720000 0.0224490000 \\
H 3.7038290000 1.9670060000 0.0214180000 \\
O 2.0453150000 2.1369090000 0.0169970000 \\
H 1.7227370000 1.6244270000 0.7990460000 \\
H 1.7263090000 1.6264310000 -0.7681660000 \\
H -1.7816160000 2.1166320000 -2.9878330000 \\
H -1.8895320000 1.9825250000 -1.4487230000 \\
O -1.7031660000 1.4830340000 -2.2695180000 \\
H 0.1603500000 1.0050480000 -2.2102460000 \\
O 1.1131330000 0.8499970000 -2.1403070000 \\
H 1.2057010000 -0.1216620000 -2.1698560000 \\
H 4.5299660000 -1.3477240000 -2.2059210000 \\
O 4.1503120000 -2.2735520000 -2.2155260000 \\
H 3.1453310000 -2.2488970000 -2.2178220000 \\
O 1.1607940000 -1.9269410000 -2.2632470000 \\
H 0.9559960000 -2.4721380000 -1.4787980000 \\
H 0.6870260000 -2.3247140000 -2.9955280000 \\
H -2.2224330000 -2.1754670000 -0.8082580000 \\
H -2.2230240000 -2.1960950000 0.7529520000 \\
O -2.0435270000 -2.7428290000 -0.0348670000 \\
H -0.5019290000 -3.2108320000 -0.0373410000 \\
O 0.4879280000 -3.3850830000 -0.0369310000 \\
H 0.6773690000 -4.3677840000 -0.0459050000 \\
H -2.1560840000 3.7162930000 0.0579700000 \\
O -2.3832480000 2.7746330000 0.0811710000 \\
H -3.3528650000 2.6797420000 0.0452800000 \\
O -4.9515700000 1.9390070000 0.0027750000 \\
H -5.4141020000 1.6125670000 0.8247850000 \\
H -5.4104170000 1.6290200000 -0.8277360000 \\
H -2.3590370000 -0.3031570000 -2.3144620000 \\
O -2.6319990000 -1.2311800000 -2.2710360000 \\
H -3.5944820000 -1.2408370000 -2.3329900000 \\
O -5.5279770000 -1.5539750000 -2.2307920000 \\
H -5.9001370000 -1.9994150000 -1.4120420000 \\
H -5.8963470000 -1.9833070000 -3.0598430000 \\
C 0.2828100000 4.4290600000 -0.0333840000 \\
H 1.1074610000 3.7075660000 -0.0169380000 \\
N -0.6608130000 5.1018440000 -0.0456510000 \\
\hline
\textbf{TS@\ce{(H2O)_{20}}} \\
\hline
H 5.0265310000 0.5452200000 1.3719930000 \\
H 5.0243990000 0.5023880000 3.0075780000 \\
O 5.2673330000 0.0037570000 2.1765170000 \\
H -1.8131370000 1.7251130000 1.5040760000 \\
H -1.8259640000 1.9623530000 3.0461000000 \\
O -1.6354100000 1.2984150000 2.3807060000 \\
H 0.1842190000 0.9593860000 2.3147530000 \\
O 1.1229330000 0.7751720000 2.1488750000 \\
H 1.1924610000 -0.1968860000 2.1648670000 \\
H 4.4646060000 -1.4572860000 2.1371270000 \\
O 4.0570330000 -2.3708920000 2.1124870000 \\
H 3.0531930000 -2.3157370000 2.1126000000 \\
O 1.0776810000 -1.9911820000 2.2040570000 \\
H 0.6083250000 -2.3981010000 2.9343450000 \\
H 0.8414370000 -2.5004300000 1.4058740000 \\
H -2.3579880000 -0.4109460000 2.3579220000 \\
O -2.6946680000 -1.3140000000 2.2391200000 \\
H -3.6568840000 -1.2510280000 2.2730780000 \\
O -5.5949530000 -1.3583940000 2.1256260000 \\
H -5.9795410000 -1.8057370000 2.9375870000 \\
H -5.9772420000 -1.7625520000 1.2902690000 \\
H 5.0327930000 0.6604830000 -3.0224560000 \\
H 5.0303750000 0.6175640000 -1.3868760000 \\
O 5.2734940000 0.1190020000 -2.2180000000 \\
H 5.1520480000 2.5119370000 0.0476240000 \\
O 4.6587140000 1.6903810000 0.0076210000 \\
H 3.7064870000 1.9218980000 0.0257500000 \\
O 1.9825860000 2.0111210000 0.0347930000 \\
H 1.6552760000 1.5542270000 0.8699970000 \\
H 1.3697740000 1.2756690000 -1.1905870000 \\
H -1.5953630000 2.0837570000 -2.7700470000 \\
H -1.7904590000 1.8992670000 -1.0877650000 \\
O -1.4594110000 1.4250110000 -2.0817720000 \\
H -0.1104420000 1.1172370000 -2.0320600000 \\
O 0.9288070000 0.8545390000 -2.0022150000 \\
H 1.0141520000 -0.1476220000 -2.0046290000 \\
H 4.4707540000 -1.3420690000 -2.2574380000 \\
O 4.0631670000 -2.2556820000 -2.2818790000 \\
H 3.0593290000 -2.2005020000 -2.2818880000 \\
O 1.0293260000 -1.7265850000 -2.1917800000 \\
H 0.8536520000 -2.3301600000 -1.4333940000 \\
H 0.4878630000 -2.0477100000 -2.9165870000 \\
H -2.3240700000 -2.0498940000 -0.8477470000 \\
H -2.3128320000 -2.1201240000 0.7227460000 \\
O -2.1198040000 -2.6303450000 -0.0893040000 \\
H -0.6127560000 -3.0873790000 -0.1115160000 \\
O 0.3716640000 -3.2916600000 -0.1154930000 \\
H 0.5307220000 -4.2793360000 -0.1411640000 \\
H -1.3593020000 3.4698450000 0.0528940000 \\
O -2.1051950000 2.4371840000 0.0136260000 \\
H -3.0653550000 2.5720510000 0.0060950000 \\
O -4.9039610000 2.1933270000 0.0209550000 \\
H -5.3755720000 1.8678210000 0.8383240000 \\
H -5.3732500000 1.9111030000 -0.8136540000 \\
H -2.3431120000 -0.1874080000 -2.2781490000 \\
O -2.6712580000 -1.1022660000 -2.3048810000 \\
H -3.6330170000 -1.0540750000 -2.3669160000 \\
O -5.5888190000 -1.2431770000 -2.2688560000 \\
H -5.9734070000 -1.6905220000 -1.4569010000 \\
H -5.9711040000 -1.6473190000 -3.1042230000 \\
C 0.6215150000 4.7154100000 0.0473350000 \\
H 1.6574640000 2.9282380000 0.0789120000 \\
N -0.4526030000 4.2449980000 0.0625480000 \\
\hline
\textbf{\ce{HNC\bond{...}(H2O)_{20}}} \\
\hline
H 5.0218320000 0.6010360000 1.3099030000 \\
H 5.0228180000 0.7043260000 2.9428170000 \\
O 5.2658820000 0.1342690000 2.1590630000 \\
H -1.8380290000 1.9453370000 1.3450460000 \\
H -1.8295400000 2.2816650000 2.8612030000 \\
O -1.6833300000 1.5693640000 2.2356170000 \\
H 0.1677080000 1.1137040000 2.2312420000 \\
O 1.1106840000 0.9373730000 2.1084190000 \\
H 1.1913520000 -0.0328660000 2.1632600000 \\
H 4.4678260000 -1.3269280000 2.2519250000 \\
O 4.0631970000 -2.2403450000 2.3097750000 \\
H 3.0591850000 -2.1884960000 2.3070780000 \\
O 1.0880520000 -1.8223370000 2.3297980000 \\
H 0.6214590000 -2.1761880000 3.0888250000 \\
H 0.8648350000 -2.4040270000 1.5766920000 \\
H -2.3859300000 -0.1869750000 2.3684880000 \\
O -2.6965950000 -1.1037830000 2.3405140000 \\
H -3.6600930000 -1.0681630000 2.3766230000 \\
O -5.5919850000 -1.2605170000 2.2528720000 \\
H -5.9736640000 -1.6347710000 3.1022570000 \\
H -5.9744820000 -1.7387310000 1.4576460000 \\
H 5.0196990000 0.3235970000 -3.0772680000 \\
H 5.0203970000 0.4268130000 -1.4443690000 \\
O 5.2636760000 -0.1431620000 -2.2282010000 \\
H 5.2077420000 2.4236420000 -0.1777200000 \\
O 4.6664640000 1.6329150000 -0.1419470000 \\
H 3.7284480000 1.9232070000 -0.1216490000 \\
O 2.0317050000 2.1490550000 -0.0473750000 \\
H 1.6932250000 1.7031330000 0.7770580000 \\
H 1.4197160000 1.1500470000 -1.4743860000 \\
H -1.5874100000 1.6987780000 -3.2766720000 \\
H -1.9367350000 2.0976490000 -1.0464860000 \\
O -1.5009920000 1.2271250000 -2.4434370000 \\
H -0.5119870000 1.0197720000 -2.3627740000 \\
O 0.9972110000 0.7043470000 -2.2299420000 \\
H 1.1304140000 -0.2582970000 -2.1209760000 \\
H 4.4655980000 -1.6043960000 -2.1353970000 \\
O 4.0609500000 -2.5177990000 -2.0773410000 \\
H 3.0569390000 -2.4659350000 -2.0801600000 \\
O 1.0880210000 -2.0472770000 -2.1080630000 \\
H 0.8724930000 -2.5515410000 -1.2985300000 \\
H 0.5994000000 -2.4656790000 -2.8190590000 \\
H -2.3151580000 -2.1769370000 -0.6592900000 \\
H -2.3079950000 -2.0905830000 0.9039010000 \\
O -2.1350820000 -2.6885980000 0.1523920000 \\
H -0.6082920000 -3.1668560000 0.1684490000 \\
O 0.3768370000 -3.3676650000 0.1806160000 \\
H 0.5389890000 -4.3531950000 0.2428730000 \\
H -1.1730920000 3.9447640000 -0.1682070000 \\
O -2.2160560000 2.6190710000 -0.2550320000 \\
H -3.1882990000 2.5569300000 -0.2483560000 \\
O -4.9163560000 2.0912660000 -0.1618660000 \\
H -5.3855210000 1.8386200000 0.6822340000 \\
H -5.3863030000 1.7343200000 -0.9670120000 \\
H -2.3454170000 -0.4561480000 -2.3227810000 \\
O -2.6914540000 -1.3526200000 -2.1880360000 \\
H -3.6522020000 -1.3016730000 -2.2529420000 \\
O -5.5942620000 -1.5379450000 -2.1343870000 \\
H -5.9758900000 -1.9122330000 -1.2849720000 \\
H -5.9767250000 -2.0161870000 -2.9295870000 \\
C 0.6214630000 5.1995220000 0.0347260000 \\
H 1.7221450000 3.0608210000 -0.0025130000 \\
N -0.3572530000 4.5670880000 -0.074612000 \\
\hline
\end{longtable}

\bibliography{MyBib.bib}